\documentclass[a4paper,10pt]{article}
\usepackage[T2A]{fontenc}
\usepackage[utf8]{inputenc}
\usepackage[english]{babel}
\usepackage{amsmath}
\usepackage{amsfonts}
\usepackage{amssymb}
\usepackage{wasysym}
\usepackage{amsthm}
\usepackage{tikz}
\usepackage{tikz-cd}
\usetikzlibrary{decorations.markings}
\usetikzlibrary{positioning}
\usepackage{braids}
\usepackage{array}
\usepackage{ytableau}
\usepackage{longtable}
\usepackage{empheq}
\usepackage{multicol}
\usepackage{mathrsfs}
\usepackage{amssymb}
\usepackage{diagbox}

\usepackage{pb-diagram}
\usepackage{setspace}
\usepackage[hidelinks,colorlinks=true,unicode]{hyperref}
\hypersetup{
linkcolor=blue,
citecolor=blue,
filecolor=magenta,
urlcolor=blue,
}
\usepackage{url}
\usepackage[left=2.2cm,right=2.2cm,top=2.2cm,bottom=2.2cm,bindingoffset=0cm]{geometry}

\begin{document}

\title{\vspace{0.1cm}{\Large {\bf A new symmetry of the colored Alexander polynomial}
\vspace{.2cm}}
\author{
{\bf V. Mishnyakov $^{a,c,d}$}\thanks{mishnyakovvv@gmail.com},
{\bf A. Sleptsov$^{a,b,c}$}\thanks{sleptsov@itep.ru}, 
{\bf N. Tselousov$^{a,c}$}\thanks{tselousov.ns@phystech.edu}} \date{ }
}
\maketitle

\vspace{-5cm}

\begin{center}
	\hfill ITEP/TH-04/20\\
	\hfill IITP/TH-04/20\\
	\hfill MIPT/TH-04/20
\end{center}

\vspace{2.3cm}

\begin{center}

$^a$ {\small {\it Institute for Theoretical and Experimental Physics, Moscow 117218, Russia}}\\
$^b$ {\small {\it Institute for Information Transmission Problems, Moscow 127994, Russia}}\\
$^c$ {\small {\it Moscow Institute of Physics and Technology, Dolgoprudny 141701, Russia }} \\
$^d$ {\small {\it Institute for Theoretical and Mathematical Physics, Moscow State University,\\ Moscow 119991, Russia }} \\

\end{center}

\vspace{1cm}

\begin{abstract}

We present a new conjectural symmetry of the colored Alexander polynomial, that is the specialization of the quantum $\mathfrak{sl}_N$ invariant widely known as the colored HOMFLY-PT polynomial. We provide arguments in support of the existence of the symmetry by studying the loop expansion and the character expansion of the colored HOMFLY-PT polynomial. We study the constraints this symmetry imposes on the group theoretic structure of the loop expansion and provide solutions to those constraints. The symmetry is a powerful tool for research on polynomial knot invariants and in the end we suggest several possible  applications of the symmetry.

\end{abstract}

\vspace{1.5cm}

\section{Introduction}
\label{inrtoduction}
The colored HOMFLY polynomial is a topological link invariant. It attracts a lot of attention because it is connected to various topics in theoretical and mathematical physics: quantum field theories \cite{Witten, ChernSimons, CS},  quantum groups \cite{RT0, RT1, RT},  conformal field theories \cite{WZNW}, topological strings \cite{TopStrings}. Whenever an explicit calculation of a class of HOMFLY invariants is derived it causes advancements in these areas. However, the computations become extremely difficult as the representation gets larger. At present moment explicit expressions are available only for several classes of knots and representations, including symmetric and anti-symmetric representations. \\
\\
In this paper we mainly discuss the symmetry of the colored Alexander polynomial $\mathcal{A}^{\mathcal{K}}_{R}(q)$ (\ref{DefAlexander}), which is a special case of colored HOMFLY polynomial. We conjecture a new "tug-the-hook" symmetry (\ref{AlexanderSym}), that relates two Alexander polynomials with different representations. One of peculiar properties of this symmetry is that it connects rectangular representations with non-rectangular ones. We also conjecture that this symmetry is valid for any knot. 
\\
\\
Let us proceed to the symmetry itself. The first nontrivial example of the action of the symmetry, that cannot be reduceв to the previously known symmetries, involves sufficiently large representations. As far as no explicit results are known for large representations except Rosso-Jones formula for torus knots \cite{RJ} we checked the symmetry only in that case. For example:
\begin{equation}
    \label{example1}
    \mathcal{A}^{3_1}_{[4,3]}(q) = \mathcal{A}^{3_1}_{[3,2,2]}(q)
\end{equation}
\begin{equation}
    \label{example2}
    \mathcal{A}^{3_1}_{[4,4]}(q) = \mathcal{A}^{3_1}_{[3,3,2]}(q)
\end{equation}
\begin{equation}
    \label{example3}
    \mathcal{A}^{3_1}_{[5,4,4]}(q) = \mathcal{A}^{3_1}_{[4,3,3]}(q)
\end{equation}
To make our statements more reasonable let us provide some arguments in support of this symmetry. These arguments are somehow knot-independent and additionally justify verifying the symmetry in the case of torus knots.
\\
\\
$\bullet$ It is well known fact that Alexander polynomial arises in 3d Chern-Simons theory with superalgebra $\mathfrak{sl}(1|1)$ \cite{alexander from supergroup 1, alexander from supergroup 2, alexander from supergroup 3, alexander from supergroup 4}. Carefull consideration of the representation theory of superalgebras reveals that Young diagrams enumerate the representations ambiguously. In general for a representation one can align several Young diagrams, that are connected via some relation. If this relation is applied to the representation of the colored Alexander polynomial $\mathcal{A}^{\mathcal{K}}_{R}(q)$ one obtains the tug-the-hook symmetry. For more careful consideration of this argument see \cite{tug the hook 2}.
\\
\\
$\bullet$ The tug-the-hook symmetry together with the rank-level duality explains vanishing of 1,3,5 orders in the loop expansion of the colored Alexander polynomial for arbitrary knot (see Section \ref{subsection vanishing of odd orders}).  This fact is surely known from the trivalent diagram point of view. All trivalent diagrams at these levels are known and we can establish the vanishing of the corresponding group-factors. However it is not obvious \emph{a priori}, without calculating the trivalent diagrams explicitly. Also we see that 2,4,6 orders of the loop expansion have the structure compatible with the symmetry. 
\\
\\
$\bullet$ The symmetry is tightly connected with the eigenvalue conjecture \cite{EC}. One of the possible formulations of the conjecture is
\textbf{the set of quantum $\mathcal{R}_i$-matrices is completely determined by the normalized eigenvalues of the quantum universal $\mathcal{\check R}$-matrix.}  The eigenvalue conjecture has been checked in numerous cases and proven in some of them (see Section 2.3 in \cite{EVind} for a review of checks of the eigenvalue conjecture). In Section \ref{section the eigenvalue conjecture} we have shown that the symmetry is a corollary of the eigenvalue conjecture.\\
It actually also strengthens our examples (\ref{example1},\ref{example2},\ref{example3}). In Section \ref{section the eigenvalue conjecture} we observe that symmetry follows from conservation of the $\mathcal{R}$ matrices eigenvalues, hence the knot itself does not appear in the proof. Hence verifying it for some knots leads to conclusion that it is true for all knots.
\\
\\
We would like to emphasize that the origin of this symmetry can be described from three different points of view: Lie superalgebras, quantum Lie algebras, and classical Lie algebras. In each case, it follows from the invariance of a certain algebraic structure.
\\
\\
The paper is organized as follows. In Section \ref{colored HOMFLY} we review some basic facts about the colored HOMFLY. In Section \ref{colored Alexander} we define the colored Alexander polynomial and discuss the motivation to study it. In Section \ref{section tug-the-hook} we present a new conjectural symmetry of the colored Alexander polynomial. We define the tug-the-hook solutions in Section \ref{section symmetry equations} to study the loop expansion. Full description of the tug-the-hook solutions is presented in Section \ref{section linear basis} in terms of combinatorics. In Section \ref{subsection derivation} we discuss derivation of the solutions. Section \ref{subsection vanishing of odd orders} is devoted to the explanation of the form of the loop expansion of the colored Alexander with the help of the symmetry. In  Section \ref{section the eigenvalue conjecture} we discuss the connection of the symmetry with the eigenvalue conjecture. The last Section \ref{discussion} is devoted to possible applications of the tug-the-hook symmetry.

\section{The colored HOMFLY polynomial}
\label{colored HOMFLY}
The HOMFLY polynomial can be arrived in various ways. We mention two of them:

     $\bullet$ \textbf{Chern-Simons theory approach.} 
     The colored HOMFLY polynomial of a knot $\mathcal{K}$ can be obtained as the average of the Wilson loop in Chern-Simons theory with the gauge group $SU(N)$ on $S^3$ \cite{Witten, Hidden}. In our notation $R$ stands for a representation of the gauge group and in particular case of $SU(N)$ is enumerated by a Young diagram.
     \begin{equation}
     \label{WilsonLoopExpValue}
         H_{R}^{\mathcal{K}}(q, a) = \left\langle \text{tr}_{R} \ P \exp \left( \oint_{\mathcal{K}} A \right) \right\rangle_{CS},
     \end{equation}
     where Chern-Simons action is given by
     \begin{equation}
         S_{CS}[A] = \frac{\kappa}{4 \pi} \int_{S^3} \text{tr} \left(  A \wedge dA +  \frac{2}{3} A \wedge A \wedge A \right).
     \end{equation}
     The polynomial variables are parameterized as follows:
     \begin{equation}
         q = e^{\hbar}, \hspace{10mm} a = e^{N \hbar}, \hspace{10mm} \hbar := \frac{2 \pi i}{\kappa + N}.
     \end{equation}
     One can evaluate $(\ref{WilsonLoopExpValue})$ in the holomorphic gauge $A_x + i A_y = 0$ \cite{La} and obtain the loop expansion of the HOMFLY polynomial $\cite{Hidden}$
     \begin{equation}
     \label{LoopExpansionHOMFLY}
         H_{R}^{\mathcal{K}}(q,a) = \sum_{n = 0}^{\infty} 
         \left( \sum_{m} v_{n, m}^{\mathcal{K}} r_{n, m}^{R} \right) \hbar^n.
     \end{equation}
     A remarkable fact about this expansion is that the knot and group dependence splits. Knot dependent parts $v^{\mathcal{K}}_{n,m}$ are integrals of fields'
     averages along the loop. They are known as Vassiliev invariants of knots or invariants of finite-type. Group dependent parts $r^{R}_{n,m}$ are called group factors and are traces of $\mathfrak{sl}_N$ generators $T_i$. 
     \begin{equation}
         r^{R}_{n,m} \sim \text{tr}_{R} \left(T_{i^{(m)}_1} T_{i^{(m)}_2}\ldots T_{i^{(m)}_{2n}} \right)
     \end{equation}
     \begin{equation}
         v^{\mathcal{K}}_{n, m} \sim \oint d x_{1} \int d x_{2} \ldots \int d x_{n}\left\langle A^{i^{(m)}_{1}}\left(x_{1}\right) A^{i^{(m)}_{2}}\left(x_2\right) \ldots A^{i^{(m)}_{n}}\left(x_{2n}\right)\right\rangle
     \end{equation}
     If one proceeds with several extra transformations of $(\ref{LoopExpansionHOMFLY})$ in the holomorphic gauge, one can arrive the Kontsevich integral form \cite{La, ChmutovDuzhin}. Before turning to this particular form let us review some basic facts about the Kontsevich integral and Lie algebra weight systems \cite{CDIntro} (see Chapter 6). Full definition of the Kontsevich integral can be found in \cite{CDIntro} (see Chapter 8.2). However, what matters for us is that its values belong to the graded completion of the algebra of unframed chord diagrams $\mathcal{\hat D}$ \cite{CDIntro} (see Chapter 4)
     \begin{equation}
         Z(\mathcal{K}) = \sum_{n = 0}^{\infty} \sum_{m} \ \mathcal{V}(\mathcal{K})_{n,m} \ \mathcal{D}_{n,m}.
     \end{equation}
     We denote $\mathcal{D}_{n,m}$ a chord diagram with $n$ chords and $\mathcal{V}(\mathcal{K})_{n,m}$ the coefficient of the chord diagram in the Kontsevich integral. Some examples of the unframed chord diagrams in small degrees are:
     
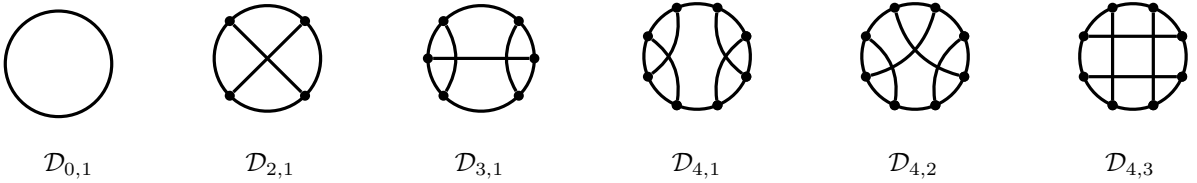
\begin{center}
     \begin{tikzpicture}
        \draw[very thick] circle (0.7cm);
    \end{tikzpicture}
    \hspace{10mm}
    \begin{tikzpicture}
        \foreach \x [count=\p] in {0,...,3} {
            \node[shape=circle,fill=black, scale=0.4] (\p) at (45-\x*90:0.7) {};
        };
        \draw[very thick] (1) arc (45:360 + 45:0.7);
        \draw[very thick] (1) to (3);
        \draw[very thick] (2) to (4);
    \end{tikzpicture}
    \hspace{10mm}
     \begin{tikzpicture}
        \foreach \x [count=\p] in {0,1,2,4,5,6} {
            \node[shape=circle,fill=black, scale=0.4] (\p) at (45-\x*45:0.7) {};
        };
        \draw[very thick] (1) arc (45:360 + 45:0.7);
        \draw[very thick] (1) to[bend right] (3);
        \draw[very thick] (4) to[bend right] (6);
        \draw[very thick] (5) to (2);
    \end{tikzpicture}
    \hspace{10mm}
     \begin{tikzpicture}
        \foreach \x [count=\p] in {0,...,11} {
            \node[shape=circle,fill=black, scale=0.4] (\p) at (45/2-\x*45:0.7) {};
        };
        \draw[very thick] (1) arc (45/2:360 + 45/2:0.7);
        \draw[very thick] (1) to[bend right] (3);
        \draw[very thick] (2) to[bend left] (8);
        \draw[very thick] (4) to[bend right] (6);
        \draw[very thick] (5) to[bend right] (7);
    \end{tikzpicture}
    \hspace{10mm}
    \begin{tikzpicture}
        \foreach \x [count=\p] in {0,...,11} {
            \node[shape=circle,fill=black, scale=0.4] (\p) at (45/2-\x*45:0.7) {};
        };
        \draw[very thick] (1) arc (45/2:360 + 45/2:0.7);
        \draw[very thick] (1) to[bend right] (3);
        \draw[very thick] (2) to[bend left] (7);
        \draw[very thick] (4) to[bend right] (6);
        \draw[very thick] (5) to[bend right] (8);
    \end{tikzpicture}
    \hspace{10mm}
     \begin{tikzpicture}
        \foreach \x [count=\p] in {0,...,11} {
            \node[shape=circle,fill=black, scale=0.4] (\p) at (45/2-\x*45:0.7) {};
        };
        \draw[very thick] (1) arc (45/2:360 + 45/2:0.7);
        \draw[very thick] (1) to (6);
        \draw[very thick] (2) to (5);
        \draw[very thick] (4) to (7);
        \draw[very thick] (3) to (8);
    \end{tikzpicture}
     \end{center}
     $$
     \mathcal{D}_{0,1} \hspace{20mm} 
     \mathcal{D}_{2,1} \hspace{21mm} 
     \mathcal{D}_{3,1} \hspace{22mm} 
     \mathcal{D}_{4,1} \hspace{22mm} 
     \mathcal{D}_{4,2} \hspace{22mm} 
     \mathcal{D}_{4,3}
     $$
     The Lie algebra weight system $\varphi_{\mathfrak{sl}_N}$ is the homomorphism from the algebra of unframed chord diagrams $\mathcal{D}$ to the center of the universal enveloping algebra $Z U(\mathfrak{sl}_N)$. The definition of $\varphi$ and the proof of this statement can be found in \cite{CDIntro} (see Chapter 6). The mapping $\varphi_{\mathfrak{sl}_N}$ is clear from examples.
     
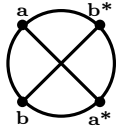
\begin{center}
        \begin{tikzpicture}[
        Circ/.style={draw,shape=circle,minimum size=14mm, node contents={},very thick}]
        \node (C2) [Circ];
        \draw[black, very thick]     (C2.north west) -- (C2.south east);
        \draw[black, very thick]     (C2.north east) -- (C2.south west);
        \fill[black]    (C2.north west) circle (2pt) node[above]{\bf{\scriptsize{a}}}
                        (C2.south west) circle (2pt) node[below]{\bf{\scriptsize{b}}}
                        (C2.south east) circle (2pt) node[below]{\bf{\scriptsize{a*}}}
                        (C2.north east) circle (2pt) node[above]{\bf{\scriptsize{b*}}};
        \node (C3) [Circ, right=63mm of C2];
        \draw[black, very thick]    (C3.north west) to[bend left] (C3.south west);
        \draw[black, very thick]    (C3.south east) to[bend left] (C3.north east);
        \draw[black, very thick]    (C3.west) -- (C3.east);
        \fill[black]    (C3.north west) circle (2pt) node[above]{\bf{\scriptsize{a}}}
                        (C3.south west) circle (2pt) node[below]{\bf{\scriptsize{a*}}}
                        (C3.south east) circle (2pt) node[below]{\bf{\scriptsize{c}}}
                        (C3.north east) circle (2pt) node[above]{\bf{\scriptsize{c*}}}
                        (C3.west) circle(2pt) node[left]{\bf{\scriptsize{b}}}
                        (C3.east) circle(2pt) node[right]{\bf{\scriptsize{b*}}}; 
        \end{tikzpicture}
     \end{center}
     $$ 
     \varphi_{\mathfrak{sl}_N}(\mathcal{D}_{2,1}) = \sum_{a,b = 1}^{\dim \mathfrak{sl}_N} T_a T_b T_a^{*} T_b^{*}
     \hspace{25mm}
     \varphi_{\mathfrak{sl}_N}(\mathcal{D}_{3,1}) = \sum_{a,b,c = 1}^{\dim \mathfrak{sl}_N} T_a T_b T_a^{*} T_c T_b^{*} T_c^{*}
     $$
     Choosing an irreducible representation of $\mathfrak{sl}_N$ identified with the Young diagram $R$ we look at the Lie algebra weight system associated with the representation $R$
     \begin{equation}
     \label{WeightSystem}
         \varphi_{\mathfrak{sl}_N}^{R} : \ \mathcal{D} \xrightarrow{\varphi_{\mathfrak{sl}_N}} Z U(\mathfrak{sl}_N) 
         \xrightarrow{\rho_R}  \text{End}(V) \xrightarrow{\text{Tr}}
         \mathbb{C}
     \end{equation}
     Linearly extending the action of $\varphi_{\mathfrak{sl}_N}^{R}$ we obtain that the loop expansion of the colored HOMFLY polynomial is the special value of the Kontsevich integral
     \begin{equation}
         \varphi_{\mathfrak{sl}_N}^{R}\left(Z(\mathcal{K}) \right) = H_{R}^{\mathcal{K}}(q,a)
     \end{equation}

\vspace{.5cm}
     
$\bullet$ \textbf{Reshetikhin-Turaev approach.} Knots and links are tightly connected to braids through the Alexander's theorem. It states that every knot or link can be represented as a closure of a braid. In this formalism \cite{RT0, RT1, RT2, LiuPeng} we construct knot invariants via representations of the braid group. The braid group $B_n$ on $n$ strands has generators $\sigma_i$, where $i = 1, \ldots, n - 1$ with the following relations on them 
\begin{center}
\begin{tabular}{l l}
    $\sigma_i \, \sigma_j = \sigma_j \, \sigma_i$ & \text{for} $|i - j| \ge 2$ \\ 
    $\sigma_i \, \sigma_{i + 1} \, \sigma_i = \sigma_{i + 1} \, \sigma_{i} \, \sigma_{i + 1}$ & \text{for} $i = 1, \ldots, n - 2$. 
\end{tabular} 
\end{center}
For example, the closure of the following braid is the trefoil knot:
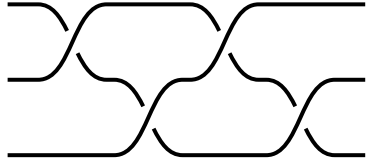
\begin{center}
\begin{tikzpicture}
    \braid[
        rotate = 90,
        number of strands = 3,
        line width = 1.5pt,
        border height =10pt,
        ]
        (braid) at (0,0)  s_2 s_1 s_2 s_1;
    \end{tikzpicture}
\end{center}
$$\sigma_2 \, \sigma_1 \, \sigma_2 \, \sigma_1 \, \in \mathcal{B}_3$$
The $i$-th strand of a $n$-strand braid is associated with the finite-dimensional module $V_i$ of the quantized universal enveloping algebra $U_q(\mathfrak{sl}_N)$ that can be fully described by the Young diagram $R_i$. It is well known \cite{KirResh, RJ} that with the help of the universal $\mathcal{\check R}$-matrix one can construct a representation of the braid group $B_n$. We consider the quantum deformation parameter $q$ that is not a root of unity:
\begin{equation}
\label{Rmat}
	\mathcal{R}_i := \pi(\sigma_i) = \mathbf{1}_1 \otimes  \ldots \otimes \mathbf{1}_{i - 1} \otimes P \mathcal{\check R} \otimes \mathbf{1}_{i + 2} \otimes \ldots \otimes \mathbf{1}_{n} \hspace{2mm} \in \hspace{2mm} \text{End}(V_{R_1} \otimes \ldots \otimes V_{R_n}).
\end{equation}
Here $P$ is the permutation operator, namely $P(x \otimes y) = y \otimes x$. The operators $\mathcal{R}_i$ satisfy the relations of the braid group $B_n$:
\begin{center}
\begin{tabular}{l l l}
    $\textbf{far commutativity property}$ & $\mathcal{R}_i \mathcal{R}_j = \mathcal{R}_j \mathcal{R}_i$ & \text{for} $|i - j| \ge 2$ \\ 
    $\textbf{braiding property}$ &  $\mathcal{R}_i \mathcal{R}_{i + 1} \mathcal{R}_i = \mathcal{R}_{i + 1} \mathcal{R}_{i} \mathcal{R}_{i + 1}$ & \text{for} $i = 1, \ldots, n - 2$. 
\end{tabular} 
\end{center}
Graphically the braiding property is the third Reidemeister move, while algebraically it is the quantum Yang-Baxter equation on the quantum $\mathcal{R}$-matrices. \\
\\
Let $\beta^{\mathcal{K}} \in B_n$ be a braid whose closure gives the knot $\mathcal{K}$. In case of knots we need only one Young diagram $R$ to describle $n$ equivalent copies of the module $V_{R}$. The closure operation corresponds to taking quantum trace and we obtain the colored HOMFLY polynomial
\begin{equation}
	H_{R}^{\mathcal{K}}(q,a) = {} _{q}\text{tr}_{V_{R}^{\otimes n}} \left( \pi( \beta^{\mathcal{K}} ) \right).
\end{equation}
For an element $z \in \text{End}(V_{R}^{\otimes n})$ the quantum trace is defined as follows:
$$_{q}\text{tr}_{V_{R}^{\otimes n}} \left(z \right) := \text{tr}_{V_{R}^{\otimes n}} \left( K_{2 \rho} \, z \right),$$
where $\rho$ is a half-sum of positive roots. In terms of simple roots $\alpha_i$ 
$$2 \rho = \sum_{i = 1}^{N - 1} n_i \, \alpha_i, \hspace{10mm} K_{2 \rho} := \prod_{i = 1}^{N - 1} K_i^{n_i}, \hspace{10mm} K_i := q^{(\alpha_i \, , \,H)}.$$
To compute the colored HOMFLY it is convinient to decompose $V_{R}^{\otimes n}$ into irreducible components:
\begin{equation}
V_{R}^{\otimes n} = \bigoplus_{Q \, \vdash n |R|} \mathcal{M}_Q \, \otimes \, V_Q.
\end{equation}
Here the sum runs over Young diagrams $Q$ that appear in the tensor product according to the Littlewood-Richardson rule. The symbol $V_Q$ stands for the module enumerated by the Young diagram $Q$, while the symbol $\mathcal{M}_Q$ stands for the space of the highest weight vector corresponding to $Q$. The dimension of the space $\mathcal{M}_Q$ is called the multiplicity of the representation $V_Q$. For example:
$$V_{[1]}^{\otimes 3} = V_{[3]} \oplus V_{[2,1]} \oplus V_{[2,1]} \oplus V_{[1,1,1]}.$$
In this case $\mathcal{M}_{[2,1]}$ is a two dimensional vector space.\\
\\
A crucial property of quantum $\mathcal{R}_i$-matrices is that they act on the modules $V_Q$ by identity:
\begin{equation}
    \mathcal{R}_i = \bigoplus_{Q \, \vdash n |R|} (\mathcal{R}_i)_{\mathcal{M}_Q} \otimes \mathbf{1}_{V_Q}.
\end{equation}
Using this fact we can simplify the expression for the colored HOMFLY. Finally it looks like a character decomposition:
\begin{equation}
\label{HOMFLYcharExpan}
    H_{R}^{\mathcal{K}} = {} _{q}\text{tr}_{V_{R}^{\otimes n}} \left( \pi( \beta^{\mathcal{K}} ) \right) = \sum_{Q \, \vdash n |R|} \text{tr}_{\, \mathcal{M}_Q} \! \left( \pi \left( \beta^{\mathcal{K}} \right) \right) \cdot \text{tr}_{\, V_Q} \! \left(K_{2\rho} \right) = 
    \sum_{Q \, \vdash n |R|} \sigma_Q \! \left( \beta^{\mathcal{K}} \right) \cdot  s_Q^{*},
\end{equation}
where $s_Q^{*}$  is a quantum dimension. The quantum dimension is defined to be Schur polynomial $s_Q$ at the special point \cite{Lin}:
\begin{equation}\label{qdim}
    s_Q^{*} :=  s_Q(x_1, \ldots, x_N) \Big|_{x_i = q^{N + 1 - 2i}}.
\end{equation}
It is worth to note that with such a definition the colored HOMFLY polynomial $H_{R}^{\mathcal{K}}$ is not actually a polynomial in the variable $q$, but a rational function.  To get a polynomial one has to normalize it by its value on the unknot, which evaluates to the quantum dimension. From now on we will work with the \textbf{normalized HOMFLY polynomial}:
\begin{equation}
\label{Normalization}
    \mathcal{H}^{\mathcal{K}}_{R} := \frac{H_{R}^{\mathcal{K}}}{H_{R}^{\bigcirc}}, \quad H_{R}^{\bigcirc}(q,a)= s_R^{*}(q,a)
\end{equation}
\vspace{0.5cm}
\\

\section{The colored Alexander polynomial}
\label{colored Alexander}
Considering the normalized colored HOMFLY polynomial we can set $a = 1$ and obtain a colored knot invariant which in the fundamental representation $R = [1]$ coincides with the famous Alexander polynomial \cite{Cromwell}. One can define the colored Alexander polynomial as follows:
\begin{equation}
\label{DefAlexander}
    \mathcal{A}^{\mathcal{K}}_{R}(q) :=  \mathcal{H}^{\mathcal{K}}_{R} (q, a = 1).
\end{equation}
The study of the colored Alexander polynomial can improve our understanding of a more complicated case of the colored HOMFLY polynomial. Our interest in the colored Alexander polynomial is also supported by the fact that it has several remarkable properties:
\begin{itemize}
    \item As a function of a representation $R$ and the quantum deformation parameter $q$ the colored Alexander polynomial respects the 1-hook scaling property:
    \begin{equation}
    \label{One hook symmetry}
    \mathcal{A}_{R}^{\mathcal{K}}(q) =  \mathcal{A}_{[1]}^{\mathcal{K}}(q^{|R|}), \hspace{10mm}
    \text{where} \hspace{10mm} R = [r, 1^L].
    \end{equation}
    This property was conjectured in \cite{IMMM} and proven for torus knots in \cite{TorusKnotProof}. Also it was shown in \cite{MM} that the eigenvalue conjecture implies this property. \\
    An example of a 1-hook Young diagram looks like:
    \begin{center}
    \ytableausetup{boxsize = 0.9em}
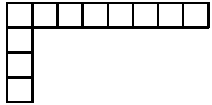
    \begin{ytableau}
    \ & \ & \ & \ & \ & \ & \ & \ \\
    \ \\
    \ \\
    \ \\
    \end{ytableau}
    \end{center}
    $$ R = [8, 1^3].$$
    \item There is the connection with the Kadomtsev-Petviashvili hierarchy \cite{KP}. The 1-hook scaling property induces so called Alexander equations \cite{MMMMS}
    \begin{equation}
         X_{n,m}([r, 1^L]) = 0.
    \end{equation} 
    These equations are homogeneous polynomials in Casimir invariants $C_1(R), C_2(R), \ldots, C_n(R)$. 
    The graded vector space of solutions of the Alexander equations, that contain monomials with even number of variables, and a ring of polynomials, generated by \textbf{the dispersion relations of 1-soliton KP $\tau$-function}, appear to be the same vector spaces \cite{MMMMS}. We provide some examples for small orders of $n$:
    \end{itemize}
    \begin{center}
    \begin{table}[h]
    \begin{doublespace}
    \begin{tabular}{|l@{\hskip 5mm}|l@{\hskip 5mm}|l|}
        \hline
        Alexander equations & Dispersion relations & KP equations \\
        \hline
        $X_{4,1} =  C_1^4 - 4 C_1 C_3 + 3 C_2^2$ 
        & $k_1^4 - 4 k_1 k_3 + 3 k_2^2 = 0$ & $\left[D_{1}^{4} - 4 D_{1} D_{3} + 3 D_{2}^{2}\right] \tau \otimes \tau=0$ \\
        $X_{5,1} = C_{2} C_{1}^{3}-3 C_{4} C_{1}+2 C_{2} C_{3}$ 
        & $k_{2} k_{1}^{3}-3 k_{4} k_{1}+2 k_{2} k_{3} = 0$ & $\left[D_{2} D_{1}^{3}  - 3 D_{4} D_{1} + 2 D_{2} D_{3}\right] \tau \otimes \tau=0 $\\
        $X_{5,2} = C_{1}\left(C_{1}^{4}-4 C_{1} C_{3}+3 C_{2}^{2}\right)$ & & \\
        \hline
    \end{tabular}
    \end{doublespace}
    \caption{Single hook solutions and relation with KP.}
    \label{AlexExamples}
    \end{table}
    \end{center}
    \vspace{-1cm}
    The dispersion relations of 1-soliton solutions can be obtained by replacing Hirota derivatives $D_i$ with $k_i$ in the KP equations in the Hirota form. Since the KP hierarchy is well studied, this correspondence gives a hope that other interesting properties of the colored Alexander polynomial can be found. \\

\section{The tug-the-hook symmetry}
\label{section tug-the-hook}
Let us present a new conjectural symmetry of the colored Alexander polynomial. To define the symmetry we introduce the Frobenius notation \cite{FrobeniusNotation} for the Young diagram $R = [R_1, R_2, \ldots, R_{l(R)}]$. This notation reflect a nice graphical interpretation of the action on Young diagrams:
\begin{equation}
\label{Frobenius}
    \alpha_i := R_i - i + 1 \hspace{10mm} \text{and} \hspace{10mm} 
    \beta_i := R^{T}_i - i + 1
\end{equation}
$$R = (\alpha_1, \ldots, \alpha_r \, | \,  \beta_1, \ldots, \beta_r)$$
where $\alpha_1 > \alpha_2 > \ldots > \alpha_r > 0$, $\beta_1 > \beta_2 > \ldots > \beta_r > 0$. \\
\\
The action on Young diagrams is defined as follows:
\begin{equation}
 \mathbf{T}_{\epsilon}(R) = (\alpha_1 + \epsilon, \dots, \alpha_r + \epsilon \, | \,  \beta_1 - \epsilon, \dots, \beta_r - \epsilon),
\end{equation}
where $\epsilon$ is an integer and $\mathbf{T}_{\epsilon}(R)$ is still a Young diagram. Note that all hooks are translated by $\epsilon$. For example:

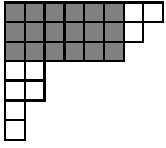
\begin{center}
\ytableausetup{boxsize = 0.7em}
\begin{multicols}{4}
        \begin{ytableau}
        *(gray) & *(gray) & *(gray) & *(gray) & *(gray) & *(gray) & \ & \ \\
        *(gray) & *(gray) & *(gray) & *(gray) & *(gray) & *(gray) & \ \\
        *(gray) & *(gray) & *(gray) & *(gray) & *(gray) & *(gray) \\
        \ & \ \\
        \ & \ \\
        \ \\
        \ \\ 
        \end{ytableau} \\
        
        
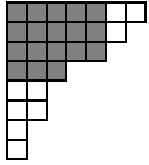
\begin{ytableau}
        *(gray) & *(gray) & *(gray) & *(gray) & *(gray) & \ & \ \\
        *(gray) & *(gray) & *(gray) & *(gray) & *(gray) & \ \\
        *(gray) & *(gray) & *(gray) & *(gray) & *(gray) \\
        *(gray) & *(gray) & *(gray) \\
        \ & \ \\
        \ & \ \\
        \ \\
        \ \\ 
        \end{ytableau} \\
       
        
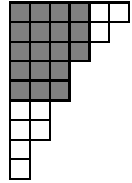
\begin{ytableau}
        *(gray) & *(gray) & *(gray) & *(gray) & \ & \ \\
        *(gray) & *(gray) & *(gray) & *(gray) & \ \\
        *(gray) & *(gray) & *(gray) & *(gray) \\
        *(gray) & *(gray) & *(gray) \\
        *(gray) & *(gray) & *(gray) \\
        \ & \ \\
        \ & \ \\
        \ \\
        \ \\ 
        \end{ytableau}
         
        
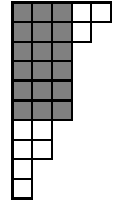
\begin{ytableau}
        *(gray) & *(gray) & *(gray) & \ & \ \\
        *(gray) & *(gray) & *(gray) & \ \\
        *(gray) & *(gray) & *(gray) \\
        *(gray) & *(gray) & *(gray) \\
        *(gray) & *(gray) & *(gray) \\
         *(gray) & *(gray) & *(gray) \\
        \ & \ \\
        \ & \ \\
        \ \\
        \ \\ 
        \end{ytableau}
\end{multicols}
\end{center}
\begin{center} 
    {\footnotesize $R = (8,6,4 \, | \, 7,4,1)$} \hspace{6mm} 
    {\footnotesize $\mathbf{T}_{-1}(R) = (7,5,3 \, | \,8,5,2)$} \hspace{6mm} 
    {\footnotesize $\mathbf{T}_{-2}(R) = (6,4,2 \, | \, 9,6,3)$} \hspace{6mm}
    {\footnotesize $\mathbf{T}_{-3}(R) = (5,3,1 \, | \, 10,7,4)$} \hspace{6mm} 
\end{center}
The tug-the-hook symmetry claims that for any possible $\epsilon$ 
\begin{equation}
\label{AlexanderSym}
    \boxed{\mathcal{A}^{\mathcal{K}}_{R}(q) = \mathcal{A}^{\mathcal{K}}_{\mathbf{T}_{\epsilon}(R)}(q)} 
\end{equation}
This new symmetry partially generalizes the 1-hook scaling property $(\ref{One hook symmetry})$. Note that the 1-hook scaling property claims that the Alexander polynomial colored with a 1-hook diagram depends only on the size of the diagram. This statement can be obtained from $(\ref{AlexanderSym})$ by substituting $R$ as a 1-hook diagram. However, the 1-hook scaling property contains additional information, namely r.h.s of $(\ref{One hook symmetry})$. The counterpart for this scaling property is not found yet for the tug-the-hook symmetry. \\
\\
It is worth to mention that the 1-hook scaling property and the tug-the-hook symmetry fails for the entire HOMFLY polynomial, for example:
\begin{equation}
    \mathcal{H}^{\mathcal{K}}_{[2]}(q,a) \not= \mathcal{H}^{\mathcal{K}}_{[1,1]}(q,a),
\end{equation}
\begin{equation}
    \mathcal{H}^{\mathcal{K}}_{[2]}(q,a) \not= \mathcal{H}^{\mathcal{K}}_{[1]}(q^2,a).
\end{equation}
It would be interesting to find the extensions of these properties to colored HOMFLY polynomials.

\section{The tug-the-hook solutions}
\label{section symmetry equations}
In this section we define the tug-the-hook solutions. If our conjecture about the tug-the-hook symmetry is true, then the group-factors should be linear combinations of the tug-the-hook solutions. The study of the basis of the tug-the-hook solutions provides us an argument in support of existence of the symmetry.
\\
\\
As the specialization of the HOMFLY invariant the Alexander polynomial inherits its perturbative expansion \eqref{LoopExpansionHOMFLY}
\begin{equation}
\label{AlexanderExpans}
\mathcal{A}_{R}^{\mathcal{K}} \left( q = e^{\hbar} \right) = \sum_{n = 0}^{\infty} \left(\sum_{m} v_{n, m}^{\mathcal{K}} r_{n, m}^{R}  \Big|_{N = 0} \right) \hbar^{n}.
\end{equation}
As we discussed in the introduction, group factors $r^{R}_{n,m}$ are the images of the Lie algebra weight system associated with the representation $R$. From the map sequence $(\ref{WeightSystem})$ it follows that $r^{R}_{n,m}$ can be expressed through the eigenvalues of the Casimir operators as they form a basis in the center of the universal enveloping algebra
\begin{equation}
\label{Group-factors}
    r_{n, m}^{R}=\sum_{|\Delta| \leq n} \alpha_{\Delta, m} C_{\Delta}(R),
\end{equation}
where we label the monomials of $C_{k}$ by the Young diagrams in accordance with the following notation:
\begin{equation}
    C_{\Delta}=\prod_{i=1}^{l(\Delta)} C_{\Delta_{i}}.
\end{equation}
Equality $(\ref{AlexanderSym})$ holds at all orders of $\hbar$ in expansion $(\ref{AlexanderExpans})$. Moreover, since Vassiliev invariants depend only on a knot, we get the following property for the Alexander group factors:
\begin{equation}
\label{condition group factors}
 r_{n, m}^{R} \Big|_{N = 0} =  r_{n, m}^{\mathbf{T}_{\epsilon}(R)} \Big|_{N = 0}.
\end{equation}

\vspace{1cm}
Now let us move to the more general problem. We consider linear combinations of monomials $C_{\Delta}$ that respect the tug-the-hook symmetry. We call these combinations \textbf{tug-the-hook solutions} $Y_{n,m}$. Indeed, group factors $r_{n, m}^{R}$ are the special case of these solutions. \\
It turns out that the Casimir invariants $C_k$ transform simply under the action of the tug-the-hook symmetry. In the next section \ref{subsection derivation} we derive explicit formulas and show that tug-the-hook solutions are homogeneous polynomials in the Casimir invariants $C_k$
\begin{equation}
\label{SymmetryEquations1}
    Y_{n,m}(R) := \sum_{|\Delta| = n} \xi^{(m)}_{\Delta} C_{\Delta}(R),
\end{equation}
\begin{equation}
\label{SymmetryEquations2}
     \boxed{ Y_{n,m}(R) = Y_{n,m}(\mathbf{T}_{\epsilon}(R))}
\end{equation}
where we enumerate by $m$ the independent solutions on the fixed level $n$. Let us denote the subspace spanned by tug-the-hook solutions  order $n$ by $\textbf{Y}_n$. Then we have $m = 1, \ldots, \dim \mathbf{Y}_n$.
\begin{equation}
    \textbf{Y}_n := \operatorname{Span} \left( \bigoplus_{m} Y_{n, m} \right)
\end{equation}
 We define a graded space of all solutions
\begin{equation}
    \textbf{Y} := \bigoplus_{n} \textbf{Y}_n
\end{equation}
In this work we are interested in the space of tug-the-hook solutions $\mathbf{Y}$. More precisely we aim to clarify two topics:
\begin{itemize}
    \item \textbf{The number of independent solutions $\dim \mathbf{Y}_n$.}
    \item \textbf{The explicit form of coefficients $\xi_{\Delta}$.}
\end{itemize}
The main motivation to study $\mathbf{Y}$ is that \textit{group-factors $r_{n,m}^{R}$ of the Alexander polynomial are linear combinations of the tug-the-hook solutions}:
\begin{equation}
r^{R}_{n, m} = \sum_{k \leq n} \sum_{l_k} \lambda_{k, l_k} Y_{k, l_k} \in \bigoplus_{k \leq n} \textbf{Y}_k
\end{equation}
The coefficients $\lambda_{k, l_k}$ are \textbf{unknown} and we leave this problem for future studies. These coefficients can be computed using some other methods and we provide an example up to the order 6 (\ref{group factors in the linear basis}).

\subsection{Linear and multiplicative basis of the tug-the-hook solutions}
\label{section linear basis}
In this section we provide the description of the explicit form the tug-the-hook solutions that we have defined in the previous section. The tug-the-hook solutions are turned to be solutions of the special linear systems of equations. These linear systems will be formulated in section \ref{subsection derivation}. In this section we also show that the construction of the full set of the tug-the-hook solution $\mathbf{Y}$ is reduced to pure combinatorics.
\\
\\
We state that the number of independent solutions on the given level $n$ is: 
\begin{equation}
\label{DimensionsY}
    \boxed{\dim \textbf{Y}_n = p(n) - p(n - 1)}
\end{equation}
where $p(n)$ is the number of Young diagrams with $n$ boxes. For small degrees it looks like
\begin{equation}
\begin{tabular}{|c|c|c|c|c|c|c|c|c|c|c|c|c|c|c|c|}
    \hline
    n & 1 & 2 & 3 & 4 & 5 & 6 & 7 & 8 & 9 & 10 & 11 & 12 & 13 & 14 & 15 \\
    \hline
    $\dim \mathbf{Y}_n$ & 1 & 1 & 1 & 2 & 2 & 4 & 4 & 7 & 8 & 12 & 14 & 21 & 24 & 34 & 41 \\
    \hline
\end{tabular}
\end{equation}
\\
Let us change the notation to simplify the formulas. From now on, we use $Y_{\Lambda}$ for $Y_{n,m}$, where $\Lambda$ is the Young diagram. Then formula $(\ref{SymmetryEquations1})$ becomes: 
\begin{equation}
\label{NewSymmetryEquation}
        Y_{\Lambda} = \sum_{|\Delta| = |\Lambda|} \xi^{\Lambda}_{\Delta} C_{\Delta}.
\end{equation}
\\
Two linear bases of $\mathbf{Y}_n$ are constructed. Here we list the properties of the first one:
\begin{enumerate}
    \item $Y_{\Lambda}$ is labeled by Young diagrams $\Lambda = [\Lambda_1, \Lambda_2, \ldots, \Lambda_r]$, where $\Lambda_1 = \Lambda_2 \geq \Lambda_3 \ldots \geq \Lambda_r$. \\
    \\
    This fact is in accordance with the formula for dimensions $(\ref{DimensionsY})$. Indeed, basis diagrams in $\mathbf{Y}_n$ do not contain diagrams that can be obtained by gluing one additional box to the first row of any diagram on the level $n - 1$. Note that $n = 1$ is the exception. \\
    \\
    We list basis diagrams up to the 8-th level:\\
    \\
    $\mathbf{Y}_1:$
    $$ \begin{ytableau}
         \  \\
         \end{ytableau} :Y_{[1]} 
    $$
    $\mathbf{Y}_2:$
    $$ \begin{ytableau}
         \ \\
         \ \\
         \end{ytableau} :Y_{[1,1]} 
    $$
    $\mathbf{Y}_3:$
    $$ \begin{ytableau}
         \ \\
         \ \\
         \ \\
         \end{ytableau} :Y_{[1,1,1]} 
    $$
    $\mathbf{Y}_4:$
    $$  \begin{ytableau}
         \ & \ \\
         \ & \ \\
         \end{ytableau} :Y_{[2,2]} 
         \hspace{10mm}
         \begin{ytableau}
         \ \\
         \ \\
         \ \\
         \ \\
         \end{ytableau} :Y_{[1,1,1,1]}
    $$
    $\mathbf{Y}_5:$
    $$  \begin{ytableau}
         \ & \ \\
         \ & \ \\
         \ \\
         \end{ytableau} :Y_{[2,2,1]} 
         \hspace{10mm}
         \begin{ytableau}
         \ \\
         \ \\
         \ \\
         \ \\
         \ \\
         \end{ytableau} :Y_{[1,1,1,1,1]}
    $$
    $\mathbf{Y}_6:$
    $$  \begin{ytableau}
         \ & \ & \ \\
         \ & \ & \ \\
         \end{ytableau} :Y_{[3,3]} 
         \hspace{10mm}
         \begin{ytableau}
         \ & \ \\
         \ & \ \\
         \ & \ \\
         \end{ytableau} :Y_{[2,2,2]}
         \hspace{10mm}
         \begin{ytableau}
         \ & \ \\
         \ & \ \\
         \ \\
         \ \\
         \end{ytableau} :Y_{[2,2,1,1]} 
         \hspace{10mm}
         \begin{ytableau}
         \ \\
         \ \\
         \ \\
         \ \\
         \ \\
         \ \\
         \end{ytableau} :Y_{[1,1,1,1,1,1]}
    $$
    $\mathbf{Y}_7:$
    $$  \begin{ytableau}
         \ & \ & \ \\
         \ & \ & \ \\
         \ \\
         \end{ytableau} :Y_{[3,3,1]} 
         \hspace{10mm}
         \begin{ytableau}
         \ & \ \\
         \ & \ \\
         \ & \ \\
         \ \\
         \end{ytableau} :Y_{[2,2,2,1]}
         \hspace{10mm}
         \begin{ytableau}
         \ & \ \\
         \ & \ \\
         \ \\
         \ \\
         \ \\
         \end{ytableau} :Y_{[2,2,1,1,1]} 
         \hspace{10mm}
         \begin{ytableau}
         \ \\
         \ \\
         \ \\
         \ \\
         \ \\
         \ \\
         \ \\
         \end{ytableau} :Y_{[1,1,1,1,1,1,1]}
    $$
    $\mathbf{Y}_8:$
    $$  
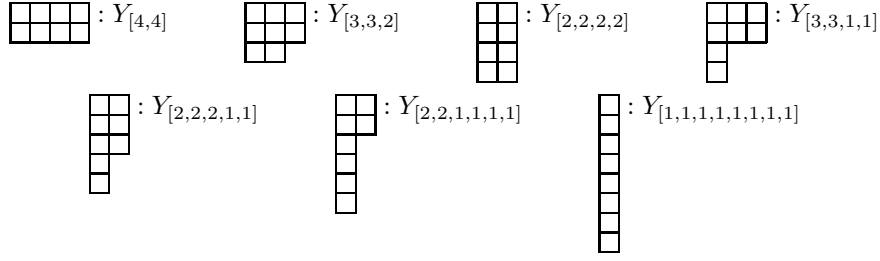
\begin{ytableau}
         \ & \ & \ & \ \\
         \ & \ & \ & \ \\
         \end{ytableau} :Y_{[4,4]} 
         \hspace{10mm}
         \begin{ytableau}
         \ & \ & \ \\
         \ & \ & \ \\
         \ & \ \\
         \end{ytableau} :Y_{[3,3,2]}
         \hspace{10mm}
         \begin{ytableau}
         \ & \ \\
         \ & \ \\
         \ & \ \\
         \ & \ \\
         \end{ytableau} :Y_{[2,2,2,2]} 
         \hspace{10mm}
         \begin{ytableau}
         \ & \ & \ \\
         \ & \ & \ \\
         \ \\
         \ \\
         \end{ytableau} :Y_{[3,3,1,1]}
    $$
    $$   \begin{ytableau}
         \ & \ \\
         \ & \ \\
         \ & \ \\
         \ \\
         \ \\
         \end{ytableau} :Y_{[2,2,2,1,1]}
         \hspace{10mm}
         \begin{ytableau}
         \ & \ \\
         \ & \ \\
         \ \\
         \ \\
         \ \\
         \ \\
         \end{ytableau} :Y_{[2,2,1,1,1,1]} 
         \hspace{10mm}
         \begin{ytableau}
         \ \\
         \ \\
         \ \\
         \ \\
         \ \\
         \ \\
         \ \\
         \ \\
         \end{ytableau} :Y_{[1,1,1,1,1,1,1,1]}
    $$
    \item The sum in formula $(\ref{NewSymmetryEquation})$ is restricted to the diagrams $\Delta$ with properties $l(\Delta) = l(\Lambda)$ and $\Delta \geq \Lambda$,  where $\geq$ means a lexicographical order.
    \begin{equation}
    \label{Result 1}
        \boxed{Y_{\Lambda} = \sum_{\substack{|\Delta| = |\Lambda| \\ l(\Delta) = l(\Lambda) \\ \Delta \geq \Lambda}} \xi_{\Delta}^{\Lambda} C_{\Delta}}
    \end{equation}
   
    \item The coefficients $\xi^{\Lambda}_{\Delta}$ have the following structure:
    \begin{equation}
    \label{Result 2}
        \boxed{\xi^{\Lambda}_{\Delta} = (-1)^{\Delta_1 - \Lambda_1} \frac{\mu^{\Lambda}_{\Delta}}{\prod_i \Delta_i !}}
    \end{equation}
    where $\mu_{\Delta}^{\Lambda}$ is the integer coefficient and its description will be given in section 5.2.
\end{enumerate}
     For illustrative purpose we list linear basis explicitly up to the 8-th level:
    \begin{center}
    \begin{doublespace}
    \begin{tabular}{l}
    \label{Linear basis1}
    $\mathbf{Y}_1:$ $Y_{[1]} = C_{[1]}$ \\
    
    $\mathbf{Y}_2:$ $Y_{[1,1]} = C_{[1,1]}$ \\
    
    $\mathbf{Y}_3:$ $Y_{[1,1,1]} = C_{[1,1,1]}$ \\
    
    $\mathbf{Y}_4:$ $Y_{[2,2]} = \frac{1}{2! 2!} C_{[2,2]} -  \frac{2}{3!} C_{[3,1]}$ \\
    $\phantom{\mathbf{Y}_4:}$ $Y_{[1,1,1,1]} = C_{[1,1,1,1]}$ \\
    
    $\mathbf{Y}_5:$ $Y_{[2,2,1]} = \frac{1}{2! 2!} C_{[2,2,1]} -  \frac{2}{3!} C_{[3,1,1]}$ \\
    $\phantom{\mathbf{Y}_5:}$ $Y_{[1,1,1,1,1]} = C_{[1,1,1,1,1]}$ \\
    
    $\mathbf{Y}_6:$ $Y_{[3,3]} = \frac{1}{3! 3!} C_{[3,3]} - \frac{2}{4! 2!} C_{[4,2]} + \frac{2}{5!} C_{[5,1]}$ \\
    $\phantom{\mathbf{Y}_6:}$ $Y_{[2,2,2]} = \frac{1}{2! 2! 2!} C_{[2,2,2]} - \frac{3}{3! 2!} C_{[3,2,1]} + \frac{3}{4!} C_{[4,1,1]}$ \\
    $\phantom{\mathbf{Y}_6:}$ $Y_{[2,2,1,1]} = \frac{1}{2! 2!} C_{[2,2,1,1]} -  \frac{2}{3!} C_{[3,1,1,1]}$ \\
    $\phantom{\mathbf{Y}_6:}$ $Y_{[1,1,1,1,1,1]} = C_{[1,1,1,1,1,1]}$ \\
    
    $\mathbf{Y}_7:$ $Y_{[3,3,1]} = \frac{1}{3! 3!} C_{[3,3,1]} - \frac{2}{4! 2!} C_{[4,2,1]} + \frac{2}{5!} C_{[5,1,1]}$ \\
    $\phantom{\mathbf{Y}_6:}$ $Y_{[2,2,2,1]} = \frac{1}{2! 2! 2!} C_{[2,2,2,1]} - \frac{3}{3! 2!} C_{[3,2,1,1]} + \frac{3}{4!} C_{[4,1,1,1]}$ \\
    $\phantom{\mathbf{Y}_6:}$ $Y_{[2,2,1,1,1]} = \frac{1}{2! 2!} C_{[2,2,1,1,1]} -  \frac{2}{3!} C_{[3,1,1,1,1]}$ \\
    $\phantom{\mathbf{Y}_6:}$ $Y_{[1,1,1,1,1,1,1]} = C_{[1,1,1,1,1,1,1]}$ \\
    $\mathbf{Y}_8:$ $Y_{[4,4]} = \frac{1}{4! 4!}C_{[4,4]} - \frac{2}{5! 3!} C_{[5,3]} + \frac{2}{6! 2!} C_{[6,2]} - \frac{2}{7!} C_{[7,1]}$ \\
    $\phantom{\mathbf{Y}_8:}$ $Y_{[3,3,2]} = \frac{1}{3! 3! 2!}C_{[3,3,2]} - \frac{2}{4! 2! 2!} C_{[4,2,2]} - \frac{1}{4!3!} C_{[4,3,1]} + \frac{5}{5! 2!} C_{[5,2,1]}  - \frac{5}{6!} C_{[6,1,1]}$ \\
    $\phantom{\mathbf{Y}_8:}$ $Y_{[3,3,1,1]} = \frac{1}{3! 3!} C_{[3,3,1,1]} - \frac{2}{4! 2!} C_{[4,2,1,1]} + \frac{2}{5!} C_{[5,1,1,1]}$ \\
    $\phantom{\mathbf{Y}_8:}$ $Y_{[2,2,2,2]} = \frac{1}{2! 2! 2! 2!}C_{[2,2,2,2]} - \frac{4}{3! 2! 2!} C_{[3,2,2,1]} + \frac{8}{4!2!} C_{[4,2,1,1]} - \frac{8}{5!} C_{[5,1,1,1]}$\\
    $\phantom{\mathbf{Y}_8:}$ $Y_{[2,2,2,1,1]} = \frac{1}{2! 2! 2!} C_{[2,2,2,1,1]} - \frac{3}{3! 2!} C_{[3,2,1,1,1]} + \frac{3}{4!} C_{[4,1,1,1,1]}$ \\
    $\phantom{\mathbf{Y}_8:}$ $Y_{[2,2,1,1,1,1]} = \frac{1}{2! 2!} C_{[2,2,1,1,1,1]} -  \frac{2}{3!} C_{[3,1,1,1,1,1]}$ \\
    $\phantom{\mathbf{Y}_8:}$ $Y_{[1,1,1,1,1,1,1,1]} = C_{[1,1,1,1,1,1,1,1]}$ \\
    \end{tabular}
    \end{doublespace}
    \end{center}
Now let us discuss the second linear basis. The set $\textbf{Y}$ has the structure of graded algebra of polynomials 
\begin{equation}
    \textbf{Y}_{n} \times \textbf{Y}_{m} \xrightarrow{} \textbf{Y}_{n + m}.
\end{equation}
We present a multiplicative basis in $\textbf{Y}$. This multiplicative basis is the subset of the first linear basis and enumerated by Young diagrams of three types
\begin{itemize}
    \item $\lambda = [\lambda_1, \lambda_2, \ldots, \lambda_r]$, where 
    $\lambda_1 = \lambda_2 > \lambda_3 > \ldots > \lambda_r \ge 2$ 
    $$
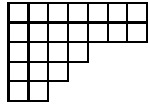
\begin{ytableau}
    \ & \ & \ & \ & \ & \ & \ \\
    \ & \ & \ & \ & \ & \ & \ \\
    \ & \ & \ & \ \\
    \ & \ & \ \\
    \ & \ \\
    \end{ytableau}$$
    \item $\lambda = [\lambda_1, \lambda_2, \ldots, \lambda_r]$, where 
    $\lambda_1 = \lambda_2 = \lambda_3 > \ldots > \lambda_r \ge 2$ 
    $$
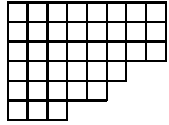
\begin{ytableau}
    \ & \ & \ & \ & \ & \ & \ & \ \\
    \ & \ & \ & \ & \ & \ & \ & \ \\
    \ & \ & \ & \ & \ & \ & \ & \ \\
     \ & \ & \ & \ & \ & \ \\
    \ & \ & \ & \ & \ \\
    \ & \ & \ \\
    \end{ytableau}$$
    \item $\lambda = [1]$
    $$\begin{ytableau}
    \ \\
    \end{ytableau}$$
\end{itemize}
Basis elements of the second linear basis are products of the multiplicative basis elements. Considering basis solutions up to the 8-th level we have the following multiplicative basis elements:
\begin{equation}
    Y_{[1]}, \hspace{6mm} Y_{[2,2]}, \hspace{6mm} Y_{[3,3]}, \hspace{6mm} Y_{[2,2,2]}, \hspace{6mm} Y_{[3,3,2]}, \hspace{6mm} Y_{[4,4]}. 
\end{equation}
Using these elements we produce the second linear basis up to the 8-th level:
\begin{center}
\begin{doublespace}
    \begin{tabular}{| c | c | c | c | c | c | c | c |}
         \hline
         $\mathbf{Y}_1$ & $\mathbf{Y}_2$ & $\mathbf{Y}_3$ & $\mathbf{Y}_4$ & $\mathbf{Y}_5$ & $\mathbf{Y}_6$ & $\mathbf{Y}_7$ & $\mathbf{Y}_8$ \\
         \hline
         $Y_{[1]}$ & $Y_{[1]}^2$ & $Y_{[1]}^3$ & $Y_{[1]}^4$ & $Y_{[1]}^5$ & $Y_{[1]}^6$ & $Y_{[1]}^7$ & $Y_{[1]}^8$ \\
         & & & $Y_{[2,2]}$ & $Y_{[2,2]}Y_{[1]}$ & $Y_{[2,2]}Y_{[1]}^2$ & $Y_{[2,2]}Y_{[1]}^3$ & $Y_{[2,2]}Y_{[1]}^4$ \\
         & & & & & $Y_{[2,2,2]}$ & $Y_{[2,2,2]}Y_{[1]}$ & $Y_{[2,2,2]}Y_{[1]}^2$ \\
         & & & & & $Y_{[3,3]}$ & $Y_{[3,3]}Y_{[1]}$ & $Y_{[3,3]}Y_{[1]}^2$ \\
         & & & & & & & $Y_{[2,2]}^2$ \\
         & & & & & & & $Y_{[3,3,2]}$ \\
         & & & & & & & $Y_{[4,4]}$ \\ [1ex]
         \hline
    \end{tabular}
\end{doublespace}
\end{center}

In this basis we observe nicely looking structure that can be generalized to an arbitrary level $n$. Considering Young diagrams that enumerate the first linear basis we "сut" them into pieces. The pieces are Young diagrams that stand for the multiplicative basis. We "cut" a diagram until it does not contain rows of equal length in the middle:
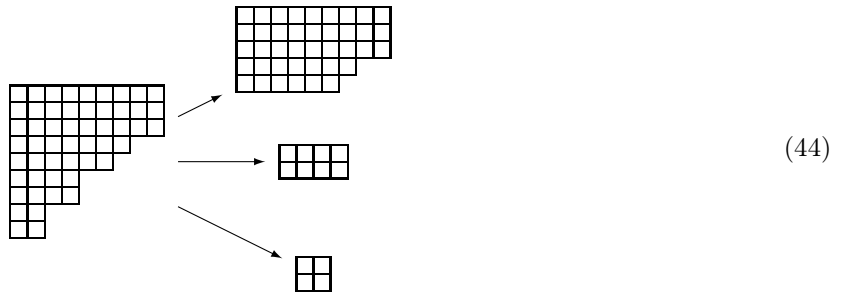
\begin{equation}
\ytableausetup{boxsize = 0.6em}
    \begin{diagram}
        \node{}
        \node{}
        \node{ \begin{ytableau}
                \ & \ & \ & \ & \ & \ & \ & \ & \ \\
                \ & \ & \ & \ & \ & \ & \ & \ & \ \\
                \ & \ & \ & \ & \ & \ & \ & \ & \ \\
                \ & \ & \ & \ & \ & \ & \ \\
                \ & \ & \ & \ & \ & \ \\
               \end{ytableau}}  \\
        \node{ \begin{ytableau}
                \ & \ & \ & \ & \ & \ & \ & \ & \ \\
                \ & \ & \ & \ & \ & \ & \ & \ & \ \\
                \ & \ & \ & \ & \ & \ & \ & \ & \ \\
                \ & \ & \ & \ & \ & \ & \ \\
                \ & \ & \ & \ & \ & \ \\
                \ & \ & \ & \ \\
                \ & \ & \ & \ \\
                \ & \ \\
                \ & \ \\
               \end{ytableau}} \arrow[2]{e,t}{} \arrow{ene,t}{} \arrow{ese}
        \node{}
        \node{ \begin{ytableau}
               \ & \ & \ & \ \\
                \ & \ & \ & \ \\
               \end{ytableau}} \\
        \node{}
        \node{}
        \node{ \begin{ytableau}
                \ & \ \\
                \ & \ \\
               \end{ytableau}}
    \end{diagram}
\end{equation}
Using formulas $(\ref{Result 1})$ and $(\ref{Result 2})$ we obtain explicit expressions for the multiplicative basis elements and thus for the linear basis element.

\subsection{Derivation of the tug-the-hook solutions}
\label{subsection derivation}
In this part of the paper we present a method to explicitly derive the tug-the-hook solutions. For this reason let us choose a basis in the the center of the universal enveloping algebra $Z U(\mathfrak{sl}_N)$ in the form:
\begin{equation}
\label{Casimirs}
    C_n(R) = \sum_{i = 1}^{l(R)} \left( R_i - i + 1 / 2 \right)^n - \left(- i + 1 / 2 \right)^n.
\end{equation}
This basis is distinguished by the following facts. The corresponding Hurwitz partition function \cite{MMN1} becomes a KP $\tau$-function  \cite{Casimir} and in terms of the Hurwitz partition function, this basis corresponds to the completed cycles and establishes a correspondence with the Gromov-Witten theory \cite{OP}. Further calculations look simplier in the Frobenius notation $(\ref{Frobenius})$. Let us denote the number of hooks in the Young diagram $R$ as $h(R)$, then
\begin{equation}
\label{CasimirsInFrobenius}
    C_n(R) = \sum_{k = 1}^{h(R)} (\alpha_k - 1/2)^n - (-\beta_k + 1/2)^n.
\end{equation}
The tug-the-hook symmetry acts on the Casimir invariants as \textbf{the translation:}
\begin{equation}
\label{Action on Casimirs}
    C_n(\mathbf{T}_{\epsilon}(R)) = \sum_i (\alpha_i + \epsilon - 1/2)^n - (-\beta_i -(-\epsilon) + 1/2)^n = \sum_{p = 0}^{n - 1} \epsilon^p \ \binom{n}{p} \ C_{n - p}(R).
\end{equation}
This formula allows us to study how monomials of the Casimir invariants transform under the action of the tug-the-hook symmetry
\begin{equation}
\label{CasimirChanging}
    C_{\Delta}(\mathbf{T}_{\epsilon}(R)) = \sum_{p = 0}^n \  \sum_{k_1 + \ldots + k_{l(\Delta)} = p} \epsilon^p \ 
    \left[ \prod_{i = 1}^{l(\Delta)} \ \binom{\Delta_i}{k_i} \ C_{\Delta_i - k_i}(R) \right].
\end{equation}
Let us consider the symmetry equation on the level $n$ $(\ref{SymmetryEquations2})$ and find constraints on coefficients $\xi_{\Delta}$. By definition tug-the-hook solutions are invariant under the action of the symmetry. Monomials of the Casimir invariants turn to the polynomials in variable $\epsilon$ $(\ref{CasimirChanging})$. We require the vanishing of coefficients in front of all positive powers of $\epsilon$. \\
By straightforward algebraic manipulations we get that the independent constraints come only from the coefficients of $\epsilon^1$ and the higher constraints are linear combinations of them. From this fact we get that the sum $(\ref{SymmetryEquations1})$ contains only Young diagrams on the level $n$. So, the coefficient of $\epsilon^{1}$ has to be zero:
\begin{equation}
\label{EquationOnCoef}
    0 = \sum_{|\Delta| = n} \xi_{\Delta} \sum_{j = 1}^{l(\Delta)} \binom{\Delta_j}{1} C_{\Delta_j - 1} \prod_{i \not= j} C_{\Delta_i}.
\end{equation}
In $(\ref{EquationOnCoef})$ the sum is over the diagrams on the $|\Delta|$-th level. This sum is actually a linear combination of the Casimir invariants on the $(|\Delta|- 1)$-th level. We represent this equation in matrix form
\begin{equation}
    \sum_{|\Delta| = n} \mathbf{M}^{(n)}_{\delta, \Delta} \xi_{\Delta} = 0,
\end{equation}
where $|\delta| = n - 1$. The matrix $\mathbf{M}^{(n)}_{\delta, \Delta}$ has two indices $\delta, \Delta$ that are Young diagrams.
Diagrams $\delta$ are ordered into sets of the increasing length $l(\delta)$. In the each set of fixed $l(\delta)$ diagrams are in the lexicographical order. Diagrams $\Delta$ are divided into two subsets. The first subset contains diagrams with unequal length of the first and second rows. The second subset contains the remaining diagrams. The diagrams in each subset are ordered in the same manner as diagrams $\delta$. Let us give an example $\mathbf{M}^{(6)}$: \\
\\
\begin{center}
\ytableausetup{boxsize = 0.4em}
    \begin{tabular}{|c||c|c|c|c|c|c|c|c|c|c|c|}
    \hline \diagbox[height=2.1cm]{\raisebox{0.5cm}{$\delta$}}{\raisebox{-0.5cm}{$\Delta$}}
    &
    \begin{ytableau}
    \ & \ & \ & \ & \ & \ \\ 
    \end{ytableau}
    & 
    \begin{ytableau}
    \ & \ & \ & \ & \ \\ 
    \ \\
    \end{ytableau}
    & 
    \begin{ytableau}
    \ & \ & \ & \ \\ 
    \ & \ \\
    \end{ytableau}
    & 
    \begin{ytableau}
    \ & \ & \ & \ \\ 
    \ \\
    \ \\
    \end{ytableau} 
    & 
    \begin{ytableau}
    \ & \ & \ \\ 
    \ & \ \\
    \ \\
    \end{ytableau} 
    &
    \begin{ytableau}
    \ & \ & \ \\ 
    \ \\
    \ \\
    \ \\
    \end{ytableau} 
    &
    \begin{ytableau}
    \ & \ \\ 
    \ \\
    \ \\
    \ \\
    \ \\
    \end{ytableau} 
    &
    \begin{ytableau}
    \ & \ & \ \\ 
    \ & \ & \ \\
    \end{ytableau} 
    &
    \begin{ytableau}
    \ & \ \\
    \ & \ \\
    \ & \ \\
    \end{ytableau} 
    &
    \begin{ytableau}
    \ & \ \\ 
    \ & \ \\
    \ \\
    \ \\
    \end{ytableau}
    &
    \begin{ytableau}
    \ \\ 
    \ \\
    \ \\
    \ \\
    \ \\ 
    \ \\ 
    \end{ytableau} 
    \\ [5ex]
    \hline
    \hline
    \begin{ytableau}
    \ & \ & \ & \ & \ \\ 
    \end{ytableau} 
    & $\raisebox{-5pt}{$\binom{6}{1}$}$ & \ & \ & \ & \ & \ & \ & \ & \ & \ & \ \\ [2ex]
    \hline
    
    \begin{ytableau}
    \ & \ & \ & \ \\ 
    \ \\
    \end{ytableau} 
    & \ & $\raisebox{-5pt}{$\binom{5}{1}$}$ & $\raisebox{-5pt}{$\binom{2}{1}$}$ & \ & \ & \ & \ & \ & \ & \ & \ \\ [2ex]
    \hline
    
    \begin{ytableau}
    \ & \ & \ \\
    \ & \ \\
    \end{ytableau} 
    & \ & \ & $\raisebox{-5pt}{$\binom{4}{1}$}$ & \ & \ & \ & \ & $\raisebox{-5pt}{$2\cdot \binom{3}{1}$}$ & \ & \ & \ \\ [2ex]
    \hline
    
    \begin{ytableau}
    \ & \ & \ \\
    \ \\
    \ \\
    \end{ytableau} 
    & \ & \ & \ & $\raisebox{-5pt}{$\binom{4}{1}$}$ & $\raisebox{-5pt}{$\binom{2}{1}$}$ & \ & \ & \ & \ & \ & \ \\ [2ex]
    \hline
    
    \begin{ytableau}
    \ & \ \\ 
    \ & \ \\
    \ \\
    \end{ytableau} 
    & \ & \ & \ & \ & $\raisebox{-7pt}{ $\binom{3}{1}$ }$ & \ & \ & \ &  $\raisebox{-7pt}{$3 \cdot \binom{2}{1}$}$ & \ & \ \\ [3ex]
    \hline
    
    \begin{ytableau}
    \ & \ \\
    \ \\
    \ \\
    \ \\
    \end{ytableau} 
    & \ & \ & \ & \ & \ & $\raisebox{-8pt}{$\binom{3}{1}$}$ & \ & \ & \ &  $\raisebox{-8pt}{$2 \cdot \binom{2}{1}$}$ & \ \\ [3ex]
    \hline
    
    \begin{ytableau}
    \ \\
    \ \\
    \ \\
    \ \\
    \ \\
    \end{ytableau} 
        & \ & \ & \ & \ & \ & \ & $\raisebox{-11pt}{$\binom{2}{1}$}$ & \ & \ & \ & \ \\ [5ex]
    \hline
\end{tabular}
\end{center}
\vspace{0.5cm}
We omit zeros in the empty boxes. In the intersection of the column $[3,3]$ and the row $[3,2]$ we have the number $2 \cdot 3$ because in sum $(\ref{EquationOnCoef})$ we have \textbf{two} nonzero terms for $j = 1$ and $j = 2$. \\
\\
We order diagrams $\delta, \Delta$ in a clever way thus the elements under the diagonal vanish. From linear algebra solutions are evident. Now one can see that diagrams with equal length of the first and the second row enumerate the solutions and we derive formula for dimensions $(\ref{DimensionsY})$.

\vspace{0.5cm}

\subsection{The description of coefficient $\mu_{\Delta}^{\Lambda}$}

In this section we provide combinatorial description of coefficient $\mu_{\Delta}^{\Lambda}$. This combinatoral algorithm arises when solving huge but sparse, up triangular linear systems of equations $\mathbf{M}^{(n)}$.
\\
\\
We construct a weighted directed graph $\Gamma_{\Lambda}$. The vertices are diagrams $\Delta$ from sum $(\ref{Result 1})$. The graph has levels from left to right according to the number of boxed in the first row, namely the diagram $\Lambda$ is at the left vertex. \\
\\
The number of outgoing edges $e$ is defined to be the number of \textbf{corner boxes.} The corner is defined as follows: it has adjacent boxes to the left and to the top, but does not have adjacent boxes to the right and to the bottom. \\
We also define a \textbf{valence} of a corner box. It is equal to the number of rows in the diagram that have the same length as the row that contains the corner box. We provide an example of a diagram to demonstrate the new definitions:

 \begin{center}
    \ytableausetup{boxsize = 1em}
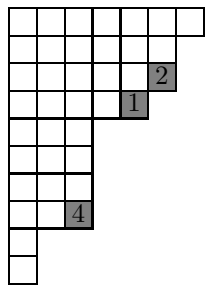
        \begin{ytableau}
        \ & \ & \ & \ & \ & \ & \ \\
        \ & \ & \ & \ & \ & \  \\
        \ & \ & \ & \ & \ & *(gray) 2 \\
        \ & \ & \ & \ & *(gray) 1 \\
        \ & \ & \\
        \ & \ & \\
        \ & \ & \\
        \ & \ & *(gray) 4 \\
        \ \\
        \ \\ 
        \end{ytableau}
\end{center}
Here we highlighted the corner boxes and put the valences into them. 
\\
\\
As mentioned above, the outgoing edges and corner boxes of a vertex-diagram are in the correspondence. An edge connects two diagrams and is weighted by the valence on the corner box $w_e$. The diagram at the head of the edge is obtained from the diagram at the tail as follows: the corner box is cut and glued to the first row of the diagram. In the examples below, notice that the rightmost diagram is a hook diagram. \\
\\
\textbf{The answer for $\mu_{\Delta}^{\Lambda}$ is given by the sum of weights over paths from $\Lambda$ to $\Delta$ in the graph  $\Gamma_{\Lambda}$:}
\begin{equation}
    \boxed{\mu_{\Delta}^{\Lambda} = \sum_{\substack{\text{paths} \\ \Lambda \rightarrow \Delta}} \ \prod_{e \, \in \, \text{path}} w_e}
\end{equation}
If the graph $\Gamma_{\Lambda}$ does not contain the vertex $\Delta$ there is no suitable path and $\mu_{\Delta}^{\Lambda} = 0$. We provide some examples:
\begin{equation}
\ytableausetup{boxsize = 0.6em}
    \begin{diagram}
        \node{ \begin{ytableau}
                \ & \ \\
                \ & *(gray)$\scriptsize{2}$ \\
               \end{ytableau}} \arrow{e,t}{2}
        \node{ \begin{ytableau}
                \ & \ & \ \\
                \ \\
               \end{ytableau}}
    \end{diagram}
\end{equation}
$$
\mu_{\text{\tiny{[3,1]}}}^{\text{\tiny{[2,2]}}} = 2
$$
\\
\begin{equation}
\ytableausetup{boxsize = 0.6em}
    \begin{diagram}
        \node{ \begin{ytableau}
                \ & \ & \ \\
                \ & \ & *(gray)$\scriptsize{2}$ \\
               \end{ytableau}} \arrow{e,t}{2}
        \node{ \begin{ytableau}
                \ & \ & \ & \ \\
                \ & *(gray)$\scriptsize{1}$  \\
               \end{ytableau}} \arrow{e,t}{1}
        \node{ \begin{ytableau}
                \ & \ & \ & \ & \ \\
                \ \\
               \end{ytableau}}
    \end{diagram}
\end{equation}
$$
\mu_{\text{\tiny{[4,2]}}}^{\text{\tiny{[3,3]}}} = 2 \hspace{10mm}
\mu_{\text{\tiny{[5,1]}}}^{\text{\tiny{[3,3]}}} = 2 \cdot 1
$$
\\
\begin{equation}
\ytableausetup{boxsize = 0.6em}
    \begin{diagram}
        \node{ \begin{ytableau}
                \ & \ \\
                \ & \ \\
                \ & *(gray)$\scriptsize{3}$ \\
               \end{ytableau}} \arrow{e,t}{3}
        \node{ \begin{ytableau}
                \ & \ & \ \\
                \ & *(gray)$\scriptsize{1}$  \\
                \ \\
               \end{ytableau}} \arrow{e,t}{1}
        \node{ \begin{ytableau}
                \ & \ & \ & \ \\
                \ \\
                \ \\
               \end{ytableau}}
    \end{diagram}
\end{equation}
$$
\mu_{\text{\tiny{[3,2,1]}}}^{\text{\tiny{[2,2,2]}}} = 3 \hspace{10mm}
\mu_{\text{\tiny{[4,1,1]}}}^{\text{\tiny{[2,2,2]}}} = 3 \cdot 1
$$
\\
\begin{equation}
\ytableausetup{boxsize = 0.6em}
    \begin{diagram}
        \node{}
        \node{ \begin{ytableau}
                \ & \ & \ & \ \\
                \ & \ \\
                \ & *(gray) $\scriptsize{2}$ \\
               \end{ytableau}} \arrow{se,t}{2} \\
        \node{ \begin{ytableau}
                \ & \ & \ \\
                \ & \ & *(gray)$\scriptsize{2}$  \\
                \ & *(gray)$\scriptsize{1}$  \\
               \end{ytableau}} \arrow{se,t}{1} \arrow{ne,t}{2}
        \node{}
        \node{ \begin{ytableau}
                \ & \ & \ & \ & \ \\
                \ & *(gray) $\scriptsize{1}$ \\
                \ \\
               \end{ytableau}} \arrow{e,t}{1} 
        \node{ \begin{ytableau}
                \ & \ & \ & \ & \ & \ \\
                \ \\
                \ \\
               \end{ytableau}} \\
        \node{}
        \node{ \begin{ytableau}
                \ & \ & \ & \ \\
                \ & \ & *(gray) $\scriptsize{1}$ \\
                \ \\
               \end{ytableau}} \arrow{ne,t}{1}
    \end{diagram}
\end{equation}
$$
\mu_{\text{\tiny{[4,2,2]}}}^{\text{\tiny{[3,3,2]}}} = 2 \hspace{10mm}
\mu_{\text{\tiny{[4,3,1]}}}^{\text{\tiny{[3,3,2]}}} = 1 \hspace{10mm}
\mu_{\text{\tiny{[5,2,1]}}}^{\text{\tiny{[3,3,2]}}} = (2 \cdot 2) + (1 \cdot 1) \hspace{10mm}
\mu_{\text{\tiny{[6,1,1]}}}^{\text{\tiny{[3,3,2]}}} = (2 \cdot 2 \cdot 1) + (1 \cdot 1 \cdot 1)
$$
\\
\begin{equation}
\ytableausetup{boxsize = 0.6em}
    \begin{diagram}
        \node{}
        \node{ \begin{ytableau}
                \ & \ & \ & \ \\
                \ & \ & *(gray) $\scriptsize{1}$ \\
                \ & \ \\
                \ & *(gray) $\scriptsize{2}$ \\
               \end{ytableau}} \arrow{e,t}{1} \arrow{sse,t}{2} 
        \node{ \begin{ytableau}
                \ & \ & \ & \ & \ \\
                \ & \ \\
                \ & \ \\
                \ & *(gray) $\scriptsize{3}$ \\
               \end{ytableau}} \arrow{e,t}{3} 
        \node{ \begin{ytableau}
                \ & \ & \ & \ & \ & \ \\
                \ & \ \\
                \ & *(gray) $\scriptsize{2}$ \\
                \ \\
               \end{ytableau}} \arrow{se,t}{2} \\
        \node{ \begin{ytableau}
                \ & \ & \ \\
                \ & \ & \ \\
                \ & \ & *(gray)$\scriptsize{3}$ \\
                \ & *(gray)$\scriptsize{1}$ \\
               \end{ytableau}}  \arrow{ne,t}{3} \arrow{se,t}{1}
        \node{}
        \node{}
        \node{}
        \node{ \begin{ytableau}
                \ & \ & \ & \ & \ & \ & \ \\
                \ & *(gray) $\scriptsize{1}$ \\
                \ \\
                \ \\
               \end{ytableau}} \arrow{e,t}{1} 
        \node{ \begin{ytableau}
                \ & \ & \ & \ & \ & \ & \ & \ \\
                \ \\
                \ \\
                \ \\
               \end{ytableau}} \\ 
        \node{}
        \node{ \begin{ytableau}
                \ & \ & \ & \ \\
                \ & \ & \ \\
                \ & \ & *(gray) $\scriptsize{2}$ \\
                \ \\
               \end{ytableau}} \arrow{e,t}{2} 
        \node{ \begin{ytableau}
                \ & \ & \ & \ & \ \\
                \ & \ & *(gray) $\scriptsize{1}$ \\
                \ & *(gray) $\scriptsize{1}$ \\
                \ \\
               \end{ytableau}} \arrow{e,t}{1} \arrow{nne,t}{1}
        \node{ \begin{ytableau}
                \ & \ & \ & \ & \ & \ \\
                \ & \ & *(gray) $\scriptsize{1}$ \\
                \ \\
                \ \\
               \end{ytableau}} \arrow{ne,t}{1}
    \end{diagram}
\end{equation}

$$
\mu_{\text{\tiny{[4,3,2,2]}}}^{\text{\tiny{[3,3,3,2]}}} = 3 \hspace{10mm}
\mu_{\text{\tiny{[4,3,3,1]}}}^{\text{\tiny{[3,3,3,2]}}} = 1 \hspace{10mm}
\mu_{\text{\tiny{[5,2,2,2]}}}^{\text{\tiny{[3,3,3,2]}}} = 3 \cdot 1 \hspace{10mm}
\mu_{\text{\tiny{[5,3,2,1]}}}^{\text{\tiny{[3,3,3,2]}}} = (3 \cdot 2)  + (1 \cdot 2) 
$$
$$
\mu_{\text{\tiny{[6,2,2,1]}}}^{\text{\tiny{[3,3,3,2]}}} = (3 \cdot 1 \cdot 3) + (3 \cdot 2 \cdot 1) + (1 \cdot 2 \cdot 1) \hspace{10mm}
\mu_{\text{\tiny{[6,3,1,1]}}}^{\text{\tiny{[3,3,3,2]}}} = (3 \cdot 2 \cdot 1) + (1 \cdot 2 \cdot 1) 
$$
$$
\mu_{\text{\tiny{[7,2,1,1]}}}^{\text{\tiny{[3,3,3,2]}}} = (3 \cdot 1 \cdot 3 \cdot 2) + (3 \cdot 2 \cdot 1 \cdot 2) + (1 \cdot 2 \cdot 1 \cdot 2) + (3 \cdot 2 \cdot 1 \cdot 1) + (1 \cdot 2 \cdot 1 \cdot 1)
$$
$$
\mu_{\text{\tiny{[8,1,1,1]}}}^{\text{\tiny{[3,3,3,2]}}} = (3 \cdot 1 \cdot 3 \cdot 2 \cdot 1) + (3 \cdot 2 \cdot 1 \cdot 2 \cdot 1) + (1 \cdot 2 \cdot 1 \cdot 2 \cdot 1) + (3 \cdot 2 \cdot 1 \cdot 1 \cdot 1) + (1 \cdot 2 \cdot 1 \cdot 1 \cdot 1)
$$

\subsection{The tug-the-hook solutions and the loop expansion}
\label{subsection vanishing of odd orders}
In this section we show that the explicit form of the tug-the-hook solutions together with the rank-level duality explains vanishing of 1,3,5 orders in the loop expansion of the colored Alexander polynomial. At the same time even orders in the expansion are linear combinations of the tug-the-hook solutions.
\\
\\
The colored Alexander polynomial respects loop expansion (\ref{AlexanderExpans}) as the special value of the colored HOMFLY. This perturbative expansion can be computed order by order with the help of trivalent diagrams calculus \cite{trivalent diagrams and the loop expansion} . These calculations become difficult very quickly, and we provide the expansion only up to the order 6:
\begin{equation}
\begin{split}
    \mathcal{A}_{R}^{\mathcal{K}} \left( q = e^{\hbar} \right) = & \, 1 + \hbar^2 \, r_{2,1}^{R} v_{2,1}^{\mathcal{K}} + \\
    & + \hbar^4 \left( r_{4,3}^{R} v_{4,3}^{\mathcal{K}} + \frac{1}{2} \left(r_{2,1}^{R} v_{2,1}^{\mathcal{K}} \right)^2 \right) \\
    & + \hbar^6 \left( r_{6,8}^{R} v_{6,8}^{\mathcal{K}} + r_{6,9}^{R} v_{6,9}^{\mathcal{K}} +
    \frac{1}{4} r_{4,3}^{R} v_{4,3}^{\mathcal{K}} r_{2,1}^{R} v_{2,1}^{\mathcal{K}} 
    + \frac{1}{6} \left(r_{2,1}^{R} v_{2,1}^{\mathcal{K}} \right)^3 
    \right) + o(\hbar^7)
\end{split}
\end{equation}
In our notation Vassiliev invariants are $v^{\mathcal{K}}_{n,m}$. The explicit form of group-factors $r^{R}_{n,m}$ reads:
\begin{equation}
\label{group factors}
    \begin{split}
        & r_{2,1}^{R} = C_1^2 \\
        & r_{4,3}^{R} = 2 \left( 3 C_2^2 - 4 C_3 C_1\right) \\
        & r_{6,8}^{R} = -\left( 12 C_5 C_1 - 30 C_4 C_2 + 20 C_3^2 \right) + 3 \left( 3 C_2^2 - 4 C_3 C_1 + C_1^4 \right) \\
        & r_{6,9}^{R} = + \left( 12 C_5 C_1 - 30 C_4 C_2 + 20 C_3^2 \right) + 5 \left( 3 C_2^2 - 4 C_3 C_1 + C_1^4 \right)
    \end{split}
\end{equation}
where $C_k$ are Casimir invariants. For simplicity we omit $R$ dependence of Casimir invariants.\\
The tug-the-hook symmetry imposes special conditions on the group-factors ($\ref{condition group factors}$). To obey these conditions a group-factor $r^{R}_{n,m}$ should be a linear combination of the tug-the-hook solutions. The linear basis (\ref{Linear basis1}) of the tug-the-hook solutions is found in this work. \\
Together with other properties of the colored Alexander polynomial this linear basis explains the form of the group-factors (\ref{group factors}). In our work we use the rank-level duality \cite{rank level duality}. The rank-level duality inherits to the colored Alexander polynomial from the colored HOMFLY polynomial:
\begin{equation}
    \mathcal{H}^{\mathcal{K}}_{R} (q, a) = \mathcal{H}^{\mathcal{K}}_{R^{T}} (q^{-1}, a) \hspace{10mm} \xrightarrow{ \ \ \ \ a = 1 \ \ \ \ } \hspace{10mm} \mathcal{A}^{\mathcal{K}}_{R} (q) = \mathcal{A}^{\mathcal{K}}_{R^{T}} (q^{-1})
\end{equation}
These properties impose the following conditions on the group-factors:
\begin{equation}
\label{rank level duality}
    r^{R^T}_{n,m} = (-1)^n \, r^{R}_{n,m}
\end{equation}
As a corollary of formula (\ref{CasimirsInFrobenius}) Casimir invariants are transforming in a simple way under the transposition of the diagram:
\begin{equation}
    C_k(R^{T}) = (-1)^{k + 1} C_k(R)
\end{equation}
Using this formula one can choose tug-the-hook solutions that obey (\ref{rank level duality}). Namely, for the particular level $n$ we consider all tug-the-hook solutions on levels $k \le n$. According to (\ref{rank level duality}) we choose tug-the-hook solutions that are multiplied by $(-1)^n$ under transposing the diagram. Let us list elements of the linear basis (\ref{Linear basis1}) that are allowed by the rank-level duality in the particular level:
\begin{center}
\begin{tabular}{c l}
1: & $\varnothing$ \\
2: & $Y_{[1,1]}$ \\
3: & $\varnothing$ \\
4: & $Y_{[2,2]}, Y_{[1,1,1,1]}$, $Y_{[1,1]}$ \\
5: & $\varnothing$ \\
6: & $Y_{[3,3]}, Y_{[2,2,1,1]}, Y_{[1,1,1,1,1,1]}$, $Y_{[2,2]}, Y_{[1,1,1,1]}$, $Y_{[1,1]}$ \\
7: & $Y_{[2,2,2,1]}$, $Y_{[2,2,2]}$
\end{tabular}
\end{center}
Finally, the actual group-factors up to the order 6 (\ref{group factors}) do not contain odd orders and even orders can be represented as a linear combination of allowed tug-the-hook solutions:
\begin{equation}
\boxed{
\label{group factors in the linear basis}
    \begin{split}
        & r_{2,1}^{R} = Y_{[1,1]} \\
        & r_{4,3}^{R} = 24 \, Y_{[2,2]} \\
        & r_{6,8}^{R} = - 720 \, Y_{[3,3]} + 36 \, Y_{[2,2]} + 3 \,  Y_{[1,1,1,1]} \\
        & r_{6,9}^{R} =  720 \, Y_{[3,3]} + 60 \, Y_{[2,2]} + 5 \,  Y_{[1,1,1,1]} \\ 
    \end{split}
    }
\end{equation}

\vspace{0.5cm}

\section{The connection with the eigenvalue conjecture}
\label{section the eigenvalue conjecture}
At this stage of the research, we do not have the complete mathematical proof that the Alexander polynomial has the tug-the-hook symmetry. Using Reshetikhin-Turaev approach, we distinguish two components in the general formula for the colored Alexander polynomial: the traces in the spaces of the highest weights and the quantum dimensions. It can be shown that the quantum dimensions are preserved under the action of the symmetry. The conservation of the traces follows from $\textbf{the eigenvalue conjecture.}$ \\
\vspace{0.5cm}

According to formula $(\ref{HOMFLYcharExpan})$ we obtain the expression for the Alexander polynomial of a $n$-strand knot colored with a representation $R$ as a sum over Young diagrams that appear in the $n$-th tensor power of $R$:
\begin{equation}
\label{AlexanderRT}
    \mathcal{A}^{\mathcal{K}}_{R}(q) = \sum_{\substack{Q \, \vdash \, n|R| \\ h(Q) = h(R)}}  \sigma_{Q}(\beta^{\mathcal{K}}) \,  \frac{s^{*}_Q}{s^{*}_R},
\end{equation}
where $h(R)$ is the number of hooks in $R$ and $s^*_R$ is the quantum dimension \eqref{qdim}. It turns out that the expansion of $\mathcal{A}^{\mathcal{K}}_{\mathbf{T}_{\epsilon}(R)}$ can be obtained by the following substitution to the formula $(\ref{AlexanderRT})$
\begin{equation}
    Q \rightarrow \mathbf{T}_{n \cdot \epsilon}(Q) \, , \hspace{5mm} R \rightarrow \mathbf{T}_{\epsilon}(R).
\end{equation}
\begin{equation}
\label{AlexanderRT2}
     \mathcal{A}^{\mathcal{K}}_{\mathbf{T}_{\epsilon}(R)}(q) = \sum_{\substack{Q \, \vdash \, n|R| \\ h(Q) = h(R)}}   \sigma_{\mathbf{T}_{n \cdot \epsilon}(Q)}(\beta^{\mathcal{K}}) \,  \frac{s^{*}_{\mathbf{T}_{n \cdot \epsilon}(Q)}}{s^{*}_{\mathbf{T}_{\epsilon}(R)}}
\end{equation}
Note that the representation $R$ is deformed with the parameter $\epsilon$ while the representation $Q$ is deformed with $n \epsilon$. 
The tug-the-hook property $(\ref{AlexanderSym})$ claims that sums $(\ref{AlexanderRT})$ and $(\ref{AlexanderRT2})$ are equal. Surprisingly, it appears that they match \textbf{pointwise:}
\begin{equation}
\label{Traces preserved}
     \sigma_{Q}(\beta^{\mathcal{K}}) = (-1)^{\epsilon \cdot (n - 1) \cdot h(R)} \ 
     \sigma_{\mathbf{T}_{n \cdot \epsilon}(Q)}(\beta^{\mathcal{K}}) 
\end{equation}
\begin{equation}
\label{Qdim preserved}
      \frac{s^{*}_Q}{s^{*}_R} \Bigg|_{N = 0} =  (-1)^{\epsilon \cdot (n - 1) \cdot h(R)}   \frac{s^{*}_{\mathbf{T}_{n \epsilon}(Q)}}{s^{*}_{\mathbf{T}_{\epsilon}(R)}} \Bigg|_{N = 0} 
\end{equation}

\vspace{1,5cm}
Note that signs in the traces and the quantum dimensions are exactly cancelled. Let us discuss some aspects of the statements above in more detail:
\begin{itemize}
\item The most simple part is the quantum dimension. The explicit formula generalizes the famous hook formula for the classical dimensions of the $SU(N)$ modules:
\begin{equation}
\label{QDIM}
    s^{*}_{Q} =  \prod_{(i, j) \in Q} \frac{[N - i + j]}{[h_{ij}]}.
\end{equation}
$$ h_{ij} := Q_i - i + Q^{\prime}_j - j + 1, \hspace{20mm} [n] := \frac{q^{n} - q^{-n}}{q - q^{-1}}.$$
In the case of the Alexander polynomial we have to take the limit $N \rightarrow 0$. In this limit all quantum dimensions are formally zero, however in $(\ref{AlexanderRT})$ we have the quantum dimension in the denominator and the ratio can be non-zero. 
It is shown in Appendix A that for arbitrary $R$ the quantum dimension can only change its sign under the action of the symmetry:
\begin{equation}
    \boxed{ \frac{s^{*}_{\mathbf{T}_{\epsilon}(R)}}{s^{*}_{R}} \Bigg|_{N = 0} = (-1)^{\epsilon \cdot h(R)}}
\end{equation}
This property implies that $(\ref{Qdim preserved})$ is satisfied.
\item
Let us recall that we denote the number of hooks by  $h(R)$. Note that $h(Q) \ge h(R)$ for $Q$ that comes from $R^{\otimes n}$ by the Littlewood-Richardson multiplication rule. From the explicit formula for the quantum dimension $(\ref{QDIM})$ in the case $N = 0$ the numerator evaluates to zero in the diagonal boxes. It means that $s^{*}_Q / s^{*}_R$ has a zero to the power of $h(Q) - h(R)$ thus the only surviving terms in the sum \eqref{AlexanderRT} are those with $h(Q) = h(R)$.
\item 
The eigenvalue conjecture was formulated in \cite{EC}. One of the possible formulations of the conjecture is
\textbf{the set of quantum $\mathcal{R}_i$-matrices is completely determined by the normalized eigenvalues of the universal $\mathcal{\check R}$-matrix.} 
\item
The eigenvalues $\lambda_W$ of the $\mathcal{R}$-matrix are defined by the irreducible representations $V_W$ that occur in the tensor square of $V_R$:
\begin{equation}
\label{DefEigen1}
V_R^{\otimes 2} = \bigoplus_{W \, \vdash 2|R|} V_{W}
\end{equation}
\begin{equation}
\label{DefEigen2}
\lambda_{W} :=  \pm q^{\varkappa (W) - 4 \varkappa(R) - N |R|} \Big|_{N = 0}
\end{equation}
\begin{equation}
\label{Kappa}
    \varkappa(W) := \sum_{(i,j) \in W} (j - i)
\end{equation}
The sign in $(\ref{DefEigen2})$ depends on whether $W$ comes from the symmetric or antisymmetric part of the tensor square. In the case of the Alexander polynomial  $N = 0$. \\

Let us denote as $\{ \lambda \}_R$ the set of eigenvalues of the $\mathcal{R}$-matrix that appears in the Alexander polynomial colored with $R$. The only important eigenvalues $\lambda_W$ are those with $h(W) = h(R)$, since in $(\ref{AlexanderRT})$ the sum runs over representations $Q$ with the constraint $h(Q) = h(R)$. 
\begin{equation}
\label{Eigenval1}
\{ \lambda \}_R = \{ \lambda_W \ | \ V_W \, \in V^{\otimes 2}_R \, , \, h(W) = h(R) \}
\end{equation}
We observe, that the set of eigenvalues in the case of $\mathbf{T}_{\epsilon}(R)$ is arranged as follows
\begin{equation}
\label{Eigenval1}
\{ \lambda \}_{\mathbf{T}_{\epsilon}(R)} = \{ \lambda_{\hat W} \ | \ V_{\hat W} \, \in V^{\otimes 2}_{\mathbf{T}_{\epsilon}(R)} \, , \, h(\hat W) = h(R) \}
\end{equation}
where
\begin{equation}
\label{W spaces}
    \hat W = {\mathbf{T}_{2 \cdot \epsilon}(W)}.
\end{equation}

From $(\ref{W spaces})$, the formula $(\ref{Action on Casimirs})$ and the fact that $\varkappa(W)$ is proportional to $C_2(W)$ we conclude that the eigenvalues are the same in the both cases
\begin{equation}
\label{EigenEquality}
\boxed{\lambda_W = \lambda_{{\mathbf{T}_{2 \cdot \epsilon}(W)}}.}
\end{equation}

\vspace{0.5cm}

Let us provide an example $\mathbf{T}_{-1}([4,3]) = [3,2,2]$. In the left column we list [4,3] case and in the right column [3,2,2].

\begin{center}
 \ytableausetup{boxsize = 0.5em}
\begin{tabular}{c c c}
     \begin{ytableau} 
     *(gray) & *(gray) & *(gray) & \ \\ *(gray) & *(gray) & *(gray) 
     \end{ytableau} & $\xrightarrow{\mathbf{T}_{-1}}$ & 
     \begin{ytableau} 
     *(gray) & *(gray) & \ \\ *(gray) & *(gray) \\ *(gray) & *(gray) 
     \end{ytableau} \\
\end{tabular}
\end{center}
\begin{center}
\begin{tabular}{l l l l l l l}
     $\lambda_{[8,6]} = q^{13}$ & $\hspace{10mm}$ &
     \begin{ytableau} 
     *(gray) & *(gray) & *(gray) & *(gray) & *(gray) & *(gray) & \ & \ \\
     *(gray) & *(gray) & *(gray) & *(gray) & *(gray) & *(gray) \\
     \end{ytableau} & $\xrightarrow{\mathbf{T}_{-2}}$ &
     \begin{ytableau} 
     *(gray) & *(gray) & *(gray) & *(gray) & \ & \ \\
     *(gray) & *(gray) & *(gray) & *(gray) \\
     *(gray) & *(gray) \\
     *(gray) & *(gray) \\
     \end{ytableau} & $\hspace{10mm}$ & $\lambda_{[6,4,2,2]} = q^{13}$ \\
     $\lambda_{[8,5,1]} = -q^{7}$ & $\hspace{10mm}$ &
     \begin{ytableau} 
     *(gray) & *(gray) & *(gray) & *(gray) & *(gray) & \ & \ & \ \\
     *(gray) & *(gray) & *(gray) & *(gray) & *(gray) \\
     \ \\
     \end{ytableau} & $\xrightarrow{\mathbf{T}_{-2}}$ &
     \begin{ytableau} 
     *(gray) & *(gray) & *(gray) & \ & \ & \ \\
     *(gray) & *(gray) & *(gray) \\
     *(gray) & *(gray) \\
     *(gray) & *(gray) \\
     \ \\
     \end{ytableau} & $\hspace{10mm}$ & $\lambda_{[6,3,2,2,1]} = -q^{7}$ \\
     $\lambda_{[8,4,2]} = q^{3}$ & $\hspace{10mm}$ &
     \begin{ytableau} 
     *(gray) & *(gray) & *(gray) & *(gray) & \ & \ & \ & \ \\
     *(gray) & *(gray) & *(gray) & *(gray) \\
     *(gray) & *(gray) \\
     \end{ytableau} & $\xrightarrow{\mathbf{T}_{-2}}$ &
     \begin{ytableau} 
     *(gray) & *(gray) & \ & \ & \ & \ \\
     *(gray) & *(gray)  \\
     *(gray) & *(gray) \\
     *(gray) & *(gray) \\
     *(gray) & *(gray) \\
     \end{ytableau} & $\hspace{10mm}$ & $\lambda_{[6,2,2,2,2]} = q^{3}$ \\
     $\lambda_{[7,7]} = -q^{11}$ & $\hspace{10mm}$ &
     \begin{ytableau} 
     *(gray) & *(gray) & *(gray) & *(gray) & *(gray) & *(gray) & *(gray)\\
     *(gray) & *(gray) & *(gray) & *(gray) & *(gray) & *(gray) & *(gray)\\
     \end{ytableau} & $\xrightarrow{\mathbf{T}_{-2}}$ &
     \begin{ytableau} 
     *(gray) & *(gray) & *(gray) & *(gray) & *(gray) \\
     *(gray) & *(gray) & *(gray) & *(gray) & *(gray) \\
     *(gray) & *(gray) \\
     *(gray) & *(gray) \\
     \end{ytableau} & $\hspace{10mm}$ & $\lambda_{[5,5,2,2]} = -q^{11}$ \\
     $\lambda_{[7,6,1]} = q^{4}$ & $\hspace{10mm}$ &
     \begin{ytableau} 
     *(gray) & *(gray) & *(gray) & *(gray) & *(gray) & *(gray) & \ \\
     *(gray) & *(gray) & *(gray) & *(gray) & *(gray) & *(gray) \\
     \ \\
     \end{ytableau} & $\xrightarrow{\mathbf{T}_{-2}}$ &
     \begin{ytableau} 
     *(gray) & *(gray) & *(gray) & *(gray) & \ \\
     *(gray) & *(gray) & *(gray) & *(gray) \\
     *(gray) & *(gray) \\
     *(gray) & *(gray) \\
     \ \\
     \end{ytableau} & $\hspace{10mm}$ & $\lambda_{[5,4,2,2,1]} = q^{4}$ \\
\end{tabular}
\end{center}
\begin{center}
\begin{tabular}{l l l l l l l}     
     $\lambda_{[7,5,2]} = q^{-1}$ & $\hspace{10mm}$ &
     \begin{ytableau} 
     *(gray) & *(gray) & *(gray) & *(gray) & *(gray) & \ & \ \\
     *(gray) & *(gray) & *(gray) & *(gray) & *(gray) \\
     *(gray) & *(gray) \\
     \end{ytableau} & $\xrightarrow{\mathbf{T}_{-2}}$ &
     \begin{ytableau} 
     *(gray) & *(gray) & *(gray) & \ & \ \\
     *(gray) & *(gray) & *(gray) \\
     *(gray) & *(gray) \\
     *(gray) & *(gray) \\
     *(gray) & *(gray) \\
     \end{ytableau} & $\hspace{10mm}$ & $\lambda_{[5,3,2,2,2]} = q^{-1}$ \\
     $\lambda_{[7,5,1,1]} = q^{-3}$ & $\hspace{10mm}$ &
     \begin{ytableau} 
     *(gray) & *(gray) & *(gray) & *(gray) & *(gray) & \ & \ \\
     *(gray) & *(gray) & *(gray) & *(gray) & *(gray) \\
     \ \\
     \ \\
     \end{ytableau} & $\xrightarrow{\mathbf{T}_{-2}}$ &
     \begin{ytableau} 
     *(gray) & *(gray) & *(gray) & \ & \ \\
     *(gray) & *(gray) & *(gray) \\
     *(gray) & *(gray) \\
     *(gray) & *(gray) \\
     \ \\
     \ \\
     \end{ytableau} & $\hspace{10mm}$ & $\lambda_{[5,3,2,2,1,1]} = q^{-3}$ \\
     $\lambda_{[7,4,2,1]} = -q^{-7}$ & $\hspace{10mm}$ &
     \begin{ytableau} 
     *(gray) & *(gray) & *(gray) & *(gray) & \ & \ & \ \\
     *(gray) & *(gray) & *(gray) & *(gray) \\
     *(gray) & *(gray) \\
     \ \\ 
     \end{ytableau} & $\xrightarrow{\mathbf{T}_{-2}}$ &
     \begin{ytableau} 
     *(gray) & *(gray) & \ & \ & \ \\
     *(gray) & *(gray) \\
     *(gray) & *(gray) \\
     *(gray) & *(gray) \\
     *(gray) & *(gray) \\
     \ \\ 
     \end{ytableau} & $\hspace{10mm}$ & $\lambda_{[5,2,2,2,2,1]} = -q^{-7}$ \\
     $\lambda_{[6,6,2]} = q^{-3}$ & $\hspace{10mm}$ &
     \begin{ytableau} 
     *(gray) & *(gray) & *(gray) & *(gray) & *(gray) & *(gray) \\
     *(gray) & *(gray) & *(gray) & *(gray) & *(gray) & *(gray) \\
     *(gray) & *(gray) \\
     \end{ytableau} & $\xrightarrow{\mathbf{T}_{-2}}$ &
     \begin{ytableau} 
     *(gray) & *(gray) & *(gray) & *(gray) \\
     *(gray) & *(gray) & *(gray) & *(gray) \\
     *(gray) & *(gray) \\
     *(gray) & *(gray) \\
     *(gray) & *(gray) \\
     \end{ytableau} & $\hspace{10mm}$ & $\lambda_{[4,4,2,2,2]} = q^{-3}$ \\
     $\lambda_{[6,6,1,1]} = -q^{-5}$ & $\hspace{10mm}$ &
     \begin{ytableau} 
     *(gray) & *(gray) & *(gray) & *(gray) & *(gray) & *(gray) \\
     *(gray) & *(gray) & *(gray) & *(gray) & *(gray) & *(gray) \\
     \ \\
     \ \\ 
     \end{ytableau} & $\xrightarrow{\mathbf{T}_{-2}}$ &
     \begin{ytableau} 
     *(gray) & *(gray) & *(gray) & *(gray) \\
     *(gray) & *(gray) & *(gray) & *(gray) \\
     *(gray) & *(gray) \\
     *(gray) & *(gray) \\
     \ \\
     \ \\ 
     \end{ytableau} & $\hspace{10mm}$ & $\lambda_{[4,4,2,2,1,1]} = -q^{-5}$ \\
     $\lambda_{[6,5,2,1]} = q^{-10}$ & $\hspace{10mm}$ &
     \begin{ytableau} 
     *(gray) & *(gray) & *(gray) & *(gray) & *(gray) & \ \\
     *(gray) & *(gray) & *(gray) & *(gray) & *(gray) \\
     *(gray) & *(gray) \\
     \ \\
     \end{ytableau} & $\xrightarrow{\mathbf{T}_{-2}}$ &
     \begin{ytableau} 
     *(gray) & *(gray) & *(gray) & \ \\
     *(gray) & *(gray) & *(gray) \\
     *(gray) & *(gray) \\
     *(gray) & *(gray) \\
     *(gray) & *(gray) \\
     \ \\
     \end{ytableau} & $\hspace{10mm}$ & $\lambda_{[4,3,2,2,2,1]} = q^{-10}$ \\
     $\lambda_{[6,4,2,2]} = q^{-15}$ & $\hspace{10mm}$ &
     \begin{ytableau} 
     *(gray) & *(gray) & *(gray) & *(gray) & \ & \ \\
     *(gray) & *(gray) & *(gray) & *(gray) \\
     *(gray) & *(gray) \\
     *(gray) & *(gray) \\
     \end{ytableau} & $\xrightarrow{\mathbf{T}_{-2}}$ &
     \begin{ytableau} 
     *(gray) & *(gray) & \ & \ \\
     *(gray) & *(gray) \\
     *(gray) & *(gray) \\
     *(gray) & *(gray) \\
     *(gray) & *(gray) \\
     *(gray) & *(gray) \\ 
     \end{ytableau} & $\hspace{10mm}$ & $\lambda_{[4,2,2,2,2,2]} = q^{-15}$ \\
     $\lambda_{[5,5,2,2]} = -q^{-17}$ & $\hspace{10mm}$ &
     \begin{ytableau} 
     *(gray) & *(gray) & *(gray) & *(gray) & *(gray) \\
     *(gray) & *(gray) & *(gray) & *(gray) & *(gray) \\
     *(gray) & *(gray) \\
     *(gray) & *(gray) \\ 
     \end{ytableau} & $\xrightarrow{\mathbf{T}_{-2}}$ &
     \begin{ytableau} 
     *(gray) & *(gray) & *(gray)  \\
     *(gray) & *(gray) & *(gray) \\
     *(gray) & *(gray) \\
     *(gray) & *(gray) \\
     *(gray) & *(gray) \\
     *(gray) & *(gray)  \\ 
     \end{ytableau} & $\hspace{10mm}$ & $\lambda_{[3,3,2,2,2,2]} = -q^{-17}$ \\
\end{tabular}
\end{center}
From $(\ref{EigenEquality})$ we conclude that the sets of eigenvalues coincide:
\begin{equation}
    \{ \lambda \}_{R} = \{ \lambda \}_{\mathbf{T}_{\epsilon}(R)}.
\end{equation}
This fact allow us to apply the eigenvalue conjecture and get $(\ref{Traces preserved})$:
\begin{equation}
    \sigma_{Q}(\beta^{\mathcal{K}})  = 
    (-1)^{\epsilon \cdot (n - 1) \cdot h(R)} \ \sigma_{\mathbf{T}_{n \cdot \epsilon}(Q)}(\beta^{\mathcal{K}}) 
\end{equation}
Note that a more careful calculation shows that the sign that appears in the traces under the action of the tug-the-hook symmetry is exactly cancelled by the sign that appear from the quantum dimensions. 
\end{itemize}

To conclude the eigenvalue conjecture one can formulate in terms of quantum 6j-symbols and their symmetries. Therefore, if this conjecture is true, then the tug-the-hook symmetry are related with some counterpart of Regge symmetries for 6j-symbols of $U_q(sl_N)$, see \cite{MS,AMS1,AMS2}. It would be very interesting to get rigorous proof of this relation in the future.

\section{Possible applications}
\label{discussion}
In conclusion let us mention several applications of the tug-the-hook symmetry. 
\begin{itemize}
    \item  \textbf{Differential expansion.} The universal differential expansion \cite{diff expansion} is yet a conjecture. However there are strong arguments in support of it. The main statement of DE is that the colored HOMFLY polynomial has the following structure:
    \begin{equation}
        \mathcal{H}^{\mathcal{K}}_R (q, a) = 1 + \sum_Q Z_{R}^{Q}(q, a) \, F^{\mathcal{K}}_Q(q, a)
    \end{equation}
    where group-theoretic and knot dependence splits. The point is that $F^{\mathcal{K}}_R(q, a)$ are much simpler than $H^{\mathcal{K}}_R(q, a)$. \\
    The tug-the-hook symmetry may be applicable to the study of the DE and give nontrivial information about $Z,F$ factors. For example, consider the following DE (for details see \cite{diff expansion}):
    \begin{equation}
    \begin{split}
        & \mathcal{H}_{[2]}^{\mathcal{K}}(q,a) = 1 + [2] F_{[1]}^{\mathcal{K}}(q, a) \{a q^2\} \{a/q\} + F_{[2]}^{\mathcal{K}}(q,a) \{a q^3 \} \{a q^2 \} \{ a / q \} \\
        & \mathcal{H}_{[1,1]}^{\mathcal{K}}(q,a) = 1 + [2] F_{[1]}^{\mathcal{K}}(q, a) \{a q\} \{a/q^2\} + F_{[1,1]}^{\mathcal{K}}(q,a) \{a q \} \{a/q^2 \} \{ a / q^3 \}
    \end{split}
    \end{equation}
    where $\{x\}  = x - x^{-1}$. From the rank-level duality we conclude $F_{[2]}^{\mathcal{K}}(q,a) = F_{[1,1]}^{\mathcal{K}}(q^{-1},a)$. If we use tug-the-hook symmetry
    \begin{equation}
        \mathcal{H}_{[2]}^{\mathcal{K}}(q,a = 1) = \mathcal{H}_{[1,1]}^{\mathcal{K}}(q,a = 1),  
    \end{equation}
    we immediately get the following
    \begin{equation}
        F_{[2]}^{\mathcal{K}}(q,a = 1) = -F_{[1,1]}^{\mathcal{K}}(q,a = 1).
    \end{equation}
    Thus, the tug-the-hook symmetry gives us additional information about the structure of the differential expansion.
    \item \textbf{Generating explicit q-holonomic relations/A-polynomials.} In papers \cite{q-holonomic} the colored HOMFLY polynomial is proven to be a $q$-holonomic function. If a function is $q$-holonomic it means existence of a recursive linear relations, also known as A-polynomials. We are not going to deep details instead we consider an explicit example from \cite{q-holonomic} for the trefoil knot:
    \begin{equation}
    \label{q holonomic}
        w_2 \mathcal{H}^{3_1}_{[m + 2]}(q, a) + w_1 \mathcal{H}^{3_1}_{[m + 1]}(q,a) + w_0 \mathcal{H}^{3_1}_{[m]}(q,a) = 0
    \end{equation}
    where $m \ge 0$ and 
    \begin{equation}
        \begin{split}
            & w_0 = a^4( a^2 q^{2m} - 1)(a^2 q^{4m + 6} - 1) \\ 
            & w_1 = q^7 \left( a^2 q^{4m + 4} - 1 \right) \left( a^4 q^{8m + 8} - a^4 q^{6m + 4} + a^4 q^{4m + 2} + a^4 q^{4m} - a^2 q^{4m + 6} - a^2 q^{4m + 2} - a^2 q^{2m} + 1 \right) \\
            & w_2 = - a^2 q^{6m + 18} \left( q^{2m + 4} - 1 \right) \left( a^2 q^{4m + 2} - 1 \right) 
        \end{split}
    \end{equation}
The general statement is well proven for any representations and any knots, however explicit examples are known only for symmetric representations as in the example above. The new symmetry allows to extend the known relations to non-symmetric representations.
\\
\\
    In this particular example we can connect three types of diagram acting with  tug-the-hook symmetry on (\ref{q holonomic}):
    \begin{equation}
       \left( w_2 \mathcal{H}^{3_1}_{[m + 2 - n_1, 1^{n_1}]} \right) \Big|_{a = 1} + \left( w_1 \mathcal{H}^{3_1}_{[m + 1 - n_2, 1^{n_2}]}\right) \Big|_{a = 1} + \left( w_0 \mathcal{H}^{3_1}_{[m - n_3, 1^{n_3}]} \right) \Big|_{a = 1} = 0
    \end{equation}
    The tug-the-hook gives a freedom in choice of $n_1, n_2, n_3$, however we lose something when setting $a = 1$. With the help of the tug-the-hook symmetry and using recursive $q$-holonomic relations one can express the colored Alexander polynomial in a particular representation via other representations. Also one can construct new recursive relations from existing ones, for example:

\begin{equation}
       \left( w_2 \mathcal{H}^{3_1}_{[m + 1, 1]} \right) \Big|_{a = 1} + \left( w_1 \mathcal{H}^{3_1}_{[m , 1]}\right) \Big|_{a = 1} + \left( w_0 \mathcal{H}^{3_1}_{[m - 1, 1]} \right) \Big|_{a = 1} = 0
\end{equation}

Note that in this example the recursion involves diagrams of the form $[m, 1]$ with one box in the second row which makes these diagrams non-rectangular.

    \item \textbf{Constructing generating functions for non-symmetric representations.} Knots-quivers correspondence attracts now more and more attention. One of the main results in that area is the following generating function for the colored HOMFLY in symmetric representations \cite{knots-quivers}:
    \begin{equation}
    \label{knot quiver generating function}
        H^{\mathcal{K}} (x) = \sum_{r = 0}^{\infty} \frac{\mathcal{H}^{\mathcal{K}}_{[r]}(q, a)}{(q^2; q^2)_r} x^r = \sum_{d_1, d_2, \ldots, d_m \ge 0}
        x^{d_1 + d_2 + \ldots + d_m} q^{\sum_{i,j} C_{ij} d_i d_j} \frac{\prod_{i=1}^{m} q^{l_i d_i} a^{a_i d_i} (-1)^{t_i d_i}}{\prod_{i=1}^{m} (q^2;q^2)_{d_i}}
    \end{equation}
    Note that $\mathcal{H}^{\mathcal{K}}_{[r]}(q, a)$ is the normalised HOMFLY polynomial and $(q^2;q^2)_r$ is the q-Pochhammer symbol. We do not discuss the right hand side of the latter formula. For the definitions and details see \cite{knots-quivers}.
    \\\\
    Finding generating functions for non-symmetric representations is an open problem.
    \\\\
    We are going to apply the tug-the-hook symmetry and invent similar generating formula for colored Alexander polynomials with 1-hook representations. For this purposes we are going to use another normalisation rule:
    \begin{equation}
        A_1^{\mathcal{K}}(x) := \sum_{r = 0}^{\infty} \mathcal{A}^{\mathcal{K}}_{[r]}(q) \, s_{[r]} (x)
    \end{equation}
    where $s_{[r]}(x)$ is one row Schur function and 1 stands for one row 1-hook diagrams in $A_1^{\mathcal{K}}$. Note that we remove the q-Pochhammer symbol. It does not change the form of (\ref{knot quiver generating function}) but should lead to change of values of $l_i, a_i, C$.
    Naively we can write the following generating function for representations of the shape $[r,1]$ with one box in the second row:
    \begin{equation}
        A^{\mathcal{K}}_{2} (x_1, x_2) = 
        \sum_{r = 1}^{\infty} \mathcal{A}^{\mathcal{K}}_{[r, 1]}(q) \, s_{[r,1]}(x_1, x_2)
    \end{equation}
    where 2 stands for two row 1-hook diagram in $A_2^{\mathcal{K}}$. Using the tug-the-hook symmetry:
    \begin{equation}
        \mathcal{A}^{\mathcal{K}}_{[r,1]}(q) = \mathcal{A}^{\mathcal{K}}_{[r + 1]}(q).
    \end{equation}
    The Schur function can be represented in the following form:
    \begin{equation}
        s_{[r,1]}(x_1, x_2) = x_1 x_2 \left( \frac{x_1^{r + 1}}{x_1 - x_2} + \frac{x_2^{r + 1}}{x_2 - x_1}\right)
    \end{equation}
    Combining the latter formulas we obtain:
    \begin{equation}
        A_2^{\mathcal{K}}(x_1, x_2) = x_1 x_2  \frac{A_1^{\mathcal{K}}(x_1) - A_1^{\mathcal{K}}(x_2)}{x_1 - x_2}
    \end{equation}
    Using more involved formula for 1-hook Schur functions we can derive generating functions with larger number of rows:
    \begin{equation}
        s_{[r, 1^{L - 1}]}(x_1, \ldots x_L) =  \prod_{k = 1}^{L} x_k \cdot \sum_{i = 1}^{L} \frac{x^{r + L - 1}_i}{\prod_{j \not = i} (x_i - x_j)}
    \end{equation}
\item \textbf{Distinguishing mutant knots.} It is known \cite{Mutant1,Mutant2,Mutant3} that HOMFLY polynomials (hence Alexander polynomials too) colored by representations associated with arbitrary rectangular Young diagram do not distinguish mutant knots. Since the tug-the-hook symmetry relates rectangular diagrams with some non-rectangular ones (see, for example, the second line in formula \eqref{example1}), then Alexander polynomials colored by such non-rectangular Young diagrams definitely do not distinguish mutant knots. 
\end{itemize}

\bigskip

Also we briefly discuss several subjects closely related to colored Alexander polynomials in which it would be interesting to find a manifestation of the tug-the-hook symmetry in the near future.
\begin{itemize}
    \item Alexander polynomials can be categorified to the knot Heegard-Floer homology \cite{HFh}. It would be interesting to promote this symmetry to the homology level. More generally, one needs to study the Khovanov-Rozansky homology and the corresponding knot polynomials. Attempts to get them via the R-matrix approach \cite{AnMor,AMP} can be very useful for this purpose.
    \item The Ooguri-Vafa partition function is a generating function for colored HOMFLY polynomials for a given knot. In the case of torus knots there is a peculiar relation of this function to a procedure of topological recursion on spectral curve \cite{BEM,DBPSS}. The tug-the-hook symmetry may provide some interesting properties of meromorphic differentials on a given spectral curve. Topological recursion, in turn, is closely related to matrix models, Virasoro algebra (and q-Virasoro \cite{LPSZ}) and integrable hierarchies. 
    \item Finally, let us mention one very promising topic, namely, a topological quantum computer. In paper \cite{KolMor} it is shown that quantum R-matrices \eqref{Rmat} are universal quantum gates. It gives an approach to the topological quantum computer, which is natural from the Chern-Simons point of view. This means that each quantum program and algorithm corresponds to some knot and polynomial knot invariant (see also \cite{MMMMM,SMtop}). Therefore, any symmetry plays an important role for the topological quantum programming. 
\end{itemize}

\section{Acknowledgements}
This work was funded by the Russian Science Foundation (Grant No.16-12-10344).

\appendix
\section{The quantum dimensions}
Here we show that the quantum dimensions can only change the sign under the action of the tug-the-hook symmetry. \\ Let us consider a Young diagram $R$ as the union of 5 parts.
\begin{enumerate}
    \item The white part of the size $h \times h$, where $h$ is the number of hooks 
    \item The green part of the size $g \times h$
    \item The yellow part of the size $y \times h$
    \item The red part that consists of $h-1$ rows of length $r_1, r_2, \dots, r_{h - 1}$
    \item The pink part that consists of $h - 1$ columns of length $p_1, p_2, \dots, p_{h - 1}$
\end{enumerate}
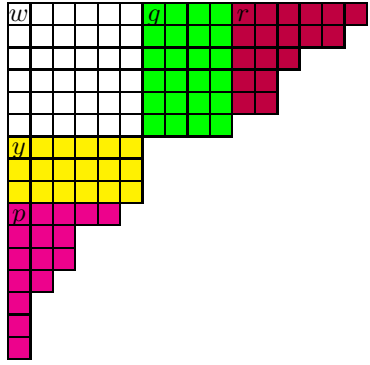
\begin{center}
\ytableausetup{boxsize = 0.8em}
        \begin{ytableau}
         *(white) w & *(white) & *(white) & *(white) & *(white) & *(white) & *(green) g & *(green) & *(green) & *(green) & *(purple) r & *(purple) & *(purple) & *(purple) & *(purple) & *(purple)\\
        *(white) & *(white) & *(white) & *(white) & *(white) & *(white) & *(green) & *(green) & *(green) & *(green) & *(purple) & *(purple) & *(purple) & *(purple) & *(purple)\\
        *(white) & *(white) & *(white) & *(white) & *(white) & *(white) & *(green) & *(green) & *(green) & *(green) & *(purple) & *(purple) & *(purple)\\
        *(white) & *(white) & *(white) & *(white) & *(white) & *(white) & *(green) & *(green) & *(green) & *(green) & *(purple) & *(purple)  \\
        *(white) & *(white) & *(white) & *(white) & *(white) & *(white) & *(green) & *(green) & *(green) & *(green) & *(purple) & *(purple)\\
        *(white) & *(white) & *(white) & *(white) & *(white) & *(white) & *(green) & *(green) & *(green) & *(green) \\
        *(yellow) y & *(yellow) & *(yellow) & *(yellow) & *(yellow) & *(yellow) \\
        *(yellow) & *(yellow) & *(yellow) & *(yellow) & *(yellow) & *(yellow) \\
        *(yellow) & *(yellow) & *(yellow) & *(yellow) & *(yellow) & *(yellow) \\
        *(magenta) p & *(magenta) & *(magenta) & *(magenta) & *(magenta) \\
        *(magenta) & *(magenta) & *(magenta) \\
        *(magenta) & *(magenta) & *(magenta) \\
        *(magenta) & *(magenta) \\
        *(magenta) \\
        *(magenta) \\
        *(magenta) \\
        \end{ytableau}
\end{center}
The formula for the quantum dimension of irreducible representation $R$
\begin{equation}
\label{AppendQDIM}
    s_R^{*} = \prod_{(i,j) \in R} \frac{[N - i + j]}{[h_{ij}]} = 
    \prod_{\text{color parts}} \ \prod_{(i,j) \in \text{part}} \frac{[N - i + j]}{[h_{ij}]}
\end{equation}
We consider the case $N = 0$. Let us note that:
\begin{enumerate}
    \item The \textbf{white} part of the product in $(\ref{AppendQDIM})$ remains the same under the action of the symmetry.
    \item The hook part (denominators) corresponding to the \textbf{red} and \textbf{pink} parts remains the same under the action of the symmetry.
\end{enumerate}
Consider the \textbf{green} and \textbf{yellow} parts entirely and the numerators in the \textbf{red} and \textbf{pink} parts. To do this consider the i-th row in the \textbf{red} part and the corresponding row in the \textbf{green} part. The contribution reads:
\begin{equation}
    \left( \frac{[r_i + g + h - i]!}{[h - i]!} \right)_{\text{red and green num.}} 
    \left( \frac{[r_i + h - i]!}{[r_i + g + h - i]!} \right)_{\text{green denom.}} = 
    \frac{[r_i + h - i]!}{[h - i]!}.
\end{equation}
We see that the contribution \textbf{does not} depend on the \textbf{green} and \textbf{yellow} parts, namely $g, b$ does not appear in the contribution. It means that this part is invariant under the tug-the-hook symmetry. \\
Next, consider the i-th column in the pink part and the corresponding column in the yellow part. The contribution reads:
\begin{equation}
\left( (-1)^{p_i + y} \ \frac{[p_i + y + h - i]!}{[h - i]!} \right)_{\text{pink and yellow num.}} 
\left( \frac{[p_i + h - i]!}{[p_i + y + h - i]!} \right)_{\text{yellow denom.}} = 
(-1)^{p_i + y} \ \frac{[p_i + h - i]!}{[h - i]!}.
\end{equation}
The contribution has $y$ dependent part $(-1)^y$. the factor $(-1)^y$ comes from each \textbf{yellow} column and we get the resulting factor $(-1)^{y h}$. Under the action of the tug-the-hook symmetry only $y, g$ parameters can change, namely $y \rightarrow y - \epsilon$ and $g \rightarrow g + \epsilon$. Considering the ratio we get 
$$ \frac{s^{*}_{\mathbf{T}_{\epsilon}(R)}}{s^{*}_{R}} \Bigg|_{N = 0} = (-1)^{\epsilon \cdot h(R)}$$
because all contributions except the factors $(-1)^{y}$ remain the same and cancel.

\end{document}